\documentclass[12pt,preprint]{emulateapj}

\usepackage{amsmath}
\usepackage{graphics}
\shorttitle{3PT With Non-linear Pressure}
\shortauthors{SHOJI \& KOMATSU}
\begin{document}
\title{
  Third-order Perturbation Theory With Non-linear Pressure
}
\author{Masatoshi Shoji \& Eiichiro Komatsu}
\affil{Texas Cosmology Center, University of Texas at Austin, \\
       1 University Station, C1400, Austin, TX, 78712}
\email{mshoji@astro.as.utexas.edu}

\begin{abstract}
We calculate the non-linear matter power spectrum using the 3rd-order
 perturbation theory without ignoring the pressure gradient
 term. We consider a semi-realistic system
 consisting of two  matter components with 
 and without  pressure, and both are expanded into the 3rd order in
 perturbations in a self-consistent manner, for the first time.
 While the pressured component may be
 identified with baryons or neutrinos, in this paper we mainly explore the
 physics of the non-linear pressure effect  using a toy model in which
 the Jeans length does not depend on time,  i.e., the sound speed
 decreases as $a^{-1/2}$, where
 $a$ is the scale factor. The linear analysis 
 shows that the power  spectrum below the so-called filtering scale is
 suppressed relative to the power spectrum of the cold dark matter. 
  Our non-linear
 calculation shows that the actual filtering scale for a given sound
 speed is smaller than the
 linear filtering scale by a factor depending on the redshift and the
 Jeans length. A $\sim 40$\% change is common, and 
 our results suggest that, when applied to baryons, the
 temperature of the 
 Inter-galactic Medium inferred from the filtering scale observed in the
 flux power spectrum of Lyman-$\alpha$ forests would be underestimated
 by a factor of two, if one used the linear filtering scale to interpret
 the data.  The 
 filtering mass, which is proportional to the filtering scale cubed, can
 also be significantly smaller than the linear theory prediction
 especially at low redshift, where the actual filtering mass can be
 smaller  than the linear prediction by a factor of three.  Finally, 
 when applied to neutrinos, we find that neutrino perturbations deviate
 significantly from linear perturbations even below the free-streaming
 scales, and thus neutrinos cannot be treated as linear perturbations.	
\end{abstract}
\section{Introduction}
Pressure plays an important role for the structure formation in the 
universe. Pressure determines the Jeans scale, $\lambda_J$, below which
the growth of structure slows down, and eventually stops and
oscillates: while fluctuations in the cold dark matter (CDM) and the
pressured component evolve in the same way above the Jeans scale, their
evolutions are significantly different below the Jeans scale.

The dominant source
of gravity is CDM, which is cold and its velocity
dispersion is negligible before the collapse of halos. However, the
sub-dominant matter components - baryons and neutrinos - have 
significant velocity dispersions, which should be included in the
calculation when precision is required. While the accurate calculations
have been done for the linear perturbations, the effects of the pressure on
the non-linear evolution of matter fluctuations on cosmological scales
($\sim 10-100$~Mpc) have not been studied very much in the literature.

We address this issue by calculating the non-linear matter power
spectrum using the 3rd-order perturbation theory \citep[3PT; see ][for a
review]{bernardeau/etal:2002}, with the
pressure gradient term in the Euler equation explicitly included. 
This enables us to study the effects of the pressure on the non-linear
evolution of matter fluctuations in a self-consistent manner.

The rest of this paper is organized as follows.
In \S~\ref{sec:baryon}, we find the linear, second-order, and
third-order solutions of the coupled continuity, Euler, and Poisson
equations for two matter components with and without the pressure
gradient.
In \S~\ref{sec:power_baryon}, we calculate the non-linear matter power
spectrum from the solutions obtained in \S~\ref{sec:baryon}.
In \S~\ref{sec:semi-analytic}, we compare our full 3PT calculation with
the approximation used by \citet{saito/etal:2008} for the effects of
massive neutrinos on the matter power spectrum.
Finally, in \S~\ref{sec:discussion}, we discuss the implications 
of our results for a few practical astrophysical and cosmological
applications. 
In Appendices we give the detailed derivations of the 3PT results used
in the main body of the paper.

\section{Third-order Perturbation Theory with Pressure}
\label{sec:baryon}
\subsection{Basic Equations}
The main goal of this paper is to find the perturbative solutions for
the CDM density contrast, $\delta_c$, for which the pressure gradient is
ignored,  and the density contrast of another matter component, $\delta_b$,
for which the pressure gradient is retained. This component may be
identified with baryons (hence the subscript ``b'') or neutrinos,
depending on the sound speed one uses in the Euler
equation.\footnote{While we use ``b'' to denote the pressured matter
component throughout this paper, we do not always mean baryons, but we 
we always refer to a general matter component with pressure.}

The equations that we are going to solve include two continuity equations:
\begin{eqnarray}
&&\dot{\delta}_c(\mathbf{x},\tau)+\mathbf{\nabla}\cdot[(1+\delta_c(\mathbf{x},\tau))\mathbf{v}_c(\mathbf{x},\tau)]=0,
\label{eq:continuity_c}\\
&&\dot{\delta}_b(\mathbf{x},\tau)+\mathbf{\nabla}\cdot[(1+\delta_b(\mathbf{x},\tau))\mathbf{v}_b(\mathbf{x},\tau)]=0,
\label{eq:continuity_b}
\end{eqnarray}
two Euler equations:
\begin{eqnarray}
&&\dot{\mathbf{v}}_c(\mathbf{x},\tau)+[\mathbf{v}_c(\mathbf{x},\tau)\cdot\mathbf{\nabla}]\mathbf{v}_c(\mathbf{x},\tau)
=-\frac{\dot{a}}{a}\mathbf{v}_c(\mathbf{x},\tau)-\mathbf{\nabla}\phi(\mathbf{x},\tau),
\label{eq:euler_c}\\
&&\dot{\mathbf{v}}_b(\mathbf{x},\tau)+[\mathbf{v}_b(\mathbf{x},\tau)\cdot\mathbf{\nabla}]\mathbf{v}_b(\mathbf{x},\tau)
=-\frac{\dot{a}}{a}\mathbf{v}_b(\mathbf{x},\tau)-\mathbf{\nabla}\phi(\mathbf{x},\tau)
\nonumber\\
&&\ \ \ \ \ \ \ \ \ \ \ \ \ \ \ \ \ \ \ \ \ \ \ \ \ \ \ \ \ \ \ \ \ \ \ 
\ \ \ \ \ \ \ \ \ \ \ 
-
\frac{c_s^2(\mathbf{x},\tau)\mathbf{\nabla}\delta_b(\mathbf{x},\tau)}{1+\delta_b(\mathbf{x},\tau)},
 \label{eq:euler_b}
\end{eqnarray}
and one Poisson equation:
\begin{eqnarray}
&&\nabla^2\phi(\mathbf{x},\tau)=4\pi Ga^2[\bar{\rho}_c(\tau)\delta_c(\mathbf{x},\tau)+\bar{\rho}_b(\tau)\delta_b(\mathbf{x},\tau)],
\end{eqnarray}
where $\delta_i\equiv (\rho_i-\bar{\rho}_i)/\bar{\rho}_i$ is the density
contrast of a matter component $i=(c,b)$, $\bar{\rho}$ the background
matter density, $a$ the scale factor, ${\mathbf v}_i$ the peculiar
velocity field  of a matter component $i$, $\phi$ the 
gravitational potential, and $c_s$ the sound speed of the matter
component with pressure. Here, the dots denote the partial derivatives
with respect to the conformal time, $\tau$, i.e., $\dot{\delta}=\partial
\delta/\partial \tau$, and $\nabla$ denotes the partial derivatives with
respect to the comoving coordinates. 

We rewrite the Poisson equation as
\begin{eqnarray}
&&\nabla^2\phi(\mathbf{x},\tau)=\frac{6}{\tau^2}\delta(\mathbf{x},\tau),
\label{eq:poisson_bc}
\end{eqnarray}
where we have assumed an Einstein-de Sitter (EdS) universe (we shall
generalize the results to other cosmological models later), for which the 
energy density of the universe is dominated entirely by the matter
density, and $a\propto \tau^2$. The background Friedmann equation
is given by 
\begin{equation}
 \frac{8\pi G}3[\bar{\rho}_c(\tau)+\bar{\rho}_b(\tau)]a^2 = \frac{4}{\tau^2}.
\end{equation}
We have also defined the total matter fluctuation, $\delta$, which is
given by 
\begin{equation}
 \delta(\mathbf{x},\tau)\equiv \frac{\bar{\rho}_c(\tau)\delta_c(\mathbf{x},\tau)+\bar{\rho}_b(\tau)\delta_b(\mathbf{x},\tau)}{\bar{\rho}_c(\tau)+\bar{\rho}_b(\tau)}=f_c\delta_c(\mathbf{x},\tau)+f_b\delta_b(\mathbf{x},\tau),
\end{equation}
where $f_c\equiv \bar{\rho}_c/(\bar{\rho}_c+\bar{\rho}_b)=\Omega_c/\Omega_m$, and
$f_b\equiv \bar{\rho}_b/(\bar{\rho}_c+\bar{\rho}_b)=\Omega_b/\Omega_m$.
For an EdS universe, $\Omega_m=1$.

Taking the divergence of the Euler equations, we obtain the equations
for the velocity divergence fields, $\theta_i\equiv \nabla\cdot {\mathbf
v}_i$. Moving non-linear terms to the right hand side (RHS)
of the equations and using the Poisson equation, we obtain
\begin{eqnarray}
&&\dot{\delta}_c(\mathbf{x},\tau)+\theta_c(\mathbf{x},\tau)
=-\mathbf\nabla\cdot[\delta_c(\mathbf{x},\tau)\mathbf{v}_c(\mathbf{x},\tau)],
\label{eq:nl_continuity_c}\\
&&\dot{\delta}_b(\mathbf{x},\tau)+\theta_b(\mathbf{x},\tau)
=-\mathbf\nabla\cdot[\delta_b(\mathbf{x},\tau)\mathbf{v}_b(\mathbf{x},\tau)],
\label{eq:nl_continuity_b}\\
&&\dot{\theta}_c(\mathbf{x},\tau)
\!+\!\frac{2}{\tau}\theta_c(\mathbf{x},\tau)\!+\!\frac{6}{\tau^2}\delta(\mathbf{x},\tau)\!
\nonumber\\
&&=\!-\mathbf{\nabla}\!\cdot\!\left\{[\mathbf{v}_c(\mathbf{x},\tau)\cdot\mathbf{\nabla}]\mathbf{v}_c(\mathbf{x},\tau)\right\},
\label{eq:nl_euler_c}\\
&&\dot{\theta}_b(\mathbf{x},\tau)
\!+\!\frac{2}{\tau}\theta_b(\mathbf{x},\tau)\!+\!\frac{6}{\tau^2}\delta(\mathbf{x},\tau)\!
\nonumber\\
&&=\!-\mathbf{\nabla}\!\cdot\!\left\{[\mathbf{v}_b(\mathbf{x},\tau)\cdot\mathbf{\nabla}]\mathbf{v}_b(\mathbf{x},\tau)\right\}
\!-\!\mathbf{\nabla}\!\cdot\!\left[\!\frac{c_s^2(\mathbf{x},\tau)\mathbf{\nabla}\delta_b(\mathbf{x},\tau)}{1+\delta_b(\mathbf{x},\tau)}\!\right].
\label{eq:nl_euler_b}
\end{eqnarray}
Note that the second term in the RHS of eq.~[\ref{eq:nl_euler_b}]
still contains the linear order term. All the other terms in the RHS of
the above equations are non-linear.

We shall simplify the pressure term, the second term in the RHS of
eq.~[\ref{eq:nl_euler_b}], as follows. First, we shall  assume that the
sound speed is homogeneous, i.e., $\nabla c_s^2=0$.
See \citet{naoz/barkana:2005} for the analysis of linear perturbations
with $\nabla c_s^2\ne 0$.
Second, we expand the pressure term to the 3rd order in perturbations:
\begin{eqnarray} 
\frac{\mathbf{\nabla}\delta\rho_b}{\rho_b}
=\frac{\mathbf{\nabla}\delta_b}{1+\delta_b}
\simeq\mathbf{\nabla}\delta_b
-\delta_b\mathbf{\nabla}\delta_b+\delta_b^2\mathbf{\nabla}\delta_b
+\mathcal{O}(\delta_b^4).
\end{eqnarray}

Going to Fourier space, we obtain
\begin{widetext}
\begin{eqnarray}
&&\dot{\tilde{\delta}}_c(\mathbf{k},\tau)+\tilde{\theta}_c(\mathbf{k},\tau)
=
-\frac1{(2\pi)^3}\int\!\int d\mathbf{q}_1d\mathbf{q}_2
\delta_D(\mathbf{q}_1+\mathbf{q}_2-\mathbf{k})
\frac{\mathbf{k}\cdot\mathbf{q}_1}{q_1^2}
\tilde{\theta}_c(\mathbf{q}_1,\tau)\tilde{\delta}_c(\mathbf{q}_2,\tau),
\label{eq:f_nl_continuity_c}\\
&&\dot{\tilde{\delta}}_b(\mathbf{k},\tau)+\tilde{\theta}_b(\mathbf{k},\tau)
=
-\frac1{(2\pi)^3}\int\!\int d\mathbf{q}_1d\mathbf{q}_2
\delta_D(\mathbf{q}_1+\mathbf{q}_2-\mathbf{k})
\frac{\mathbf{k}\cdot\mathbf{q}_1}{q_1^2}
\tilde{\theta}_b(\mathbf{q}_1,\tau)\tilde{\delta}_b(\mathbf{q}_2,\tau),
\label{eq:f_nl_continuity_b}\\
&&\dot{\tilde{\theta}}_c(\mathbf{k},\tau)+
\frac{2}{\tau}\tilde{\theta}_c(\mathbf{k},\tau)+\frac{6}{\tau^2}
\tilde{\delta}(\mathbf{k},\tau)
=
-\frac1{(2\pi)^3}\int\!\int d\mathbf{q}_1d\mathbf{q}_2
\delta_D(\mathbf{q}_1+\mathbf{q}_2-\mathbf{k})
\frac{k^2(\mathbf{q}_1\cdot\mathbf{q}_2)}{2q_1^2q_2^2}
\tilde{\theta}_c(\mathbf{q}_1,\tau)\tilde{\theta}_c(\mathbf{q}_2,\tau),
\label{eq:f_nl_euler_c}\\
&&\dot{\tilde{\theta}}_b(\mathbf{k},\tau)
+\frac{2}{\tau}\tilde{\theta}_b(\mathbf{k},\tau)
+\frac{6}{\tau^2}\tilde{\delta}(\mathbf{k},\tau)
=
-\frac1{(2\pi)^3}\int\!\int d\mathbf{q}_1d\mathbf{q}_2
\delta_D(\mathbf{q}_1+\mathbf{q}_2-\mathbf{k})
\frac{k^2(\mathbf{q}_1\cdot\mathbf{q}_2)}{2q_1^2q_2^2}
\tilde{\theta}_b(\mathbf{q}_1,\tau)\tilde{\theta}_b(\mathbf{q}_2,\tau)
\nonumber\\
&&\ \ \ \ \ \ \ \ \ \ \ \ \ \ \ \ \ \ \ \ \ \ \ \ \ \ \ \ \ \ \ \ \ \ \ \ \ \ \ \ \ \ \ \ \ -\mathcal{F}
\left[\mathbf{\nabla}\cdot\left(\frac{c_s^2(\tau)\mathbf{\nabla}\delta_b(\mathbf{x},\tau)}{1+\delta_b(\mathbf{x},\tau)}\right)\right](\mathbf{k}),
\label{eq:f_nl_euler_b}
\end{eqnarray}
where
\begin{eqnarray} 
\mathcal{F}\left[\mathbf{\nabla}\cdot\left(\frac{c_s^2(\tau)\mathbf{\nabla}\delta_b(\mathbf{x},\tau)}{1+\delta_b(\mathbf{x},\tau)}\right)
\right](\mathbf{k})
&&\equiv -k^2c_s^2(\tau)\left[
\tilde{\delta}_b(\mathbf{k})
-\frac1{2(2\pi)^3}\int\!\int d\mathbf{q}_1d\mathbf{q}_2
\tilde{\delta}_b(\mathbf{q}_1,\tau)\tilde{\delta}_b(\mathbf{q}_2,\tau)
\delta_D(\mathbf{q}_1+\mathbf{q}_2-\mathbf{k})
\right.
\nonumber\\
&+&\frac1{3(2\pi)^6}\left.\int\!\int\!\int d\mathbf{q}_1
d\mathbf{q}_2d\mathbf{q}_3
\tilde{\delta}_b(\mathbf{q}_1,\tau)\tilde{\delta}_b(\mathbf{q}_2,\tau)
\tilde{\delta}_b(\mathbf{q}_3,\tau)
\delta_D(\mathbf{q}_1+\mathbf{q}_2+\mathbf{q}_3-\mathbf{k})
\right].
\end{eqnarray}
\end{widetext}

In the subsequent subsections we shall solve these coupled equations
perturbatively. Hereafter we shall omit the tildes on the perturbation
variables in Fourier space.
\subsection{Linear Order Solution: Jeans Filtering Scale}

In the linear order, one finds
\begin{eqnarray}
\label{eq:cont_linear}
&&\dot{\delta}_{1,c}(\mathbf{k},\tau)+\theta_{1,c}(\mathbf{k},\tau)=0,\\
&&\dot{\delta}_{1,b}(\mathbf{k},\tau)+\theta_{1,b}(\mathbf{k},\tau)=0,\\
&&\dot{\theta}_{1,c}(\mathbf{k},\tau)+\frac{2}{\tau}\theta_{1,c}(\mathbf{k},\tau)+\frac{6}{\tau^2}
\delta_1(\mathbf{k},\tau)=0,\\
&&\dot{\theta}_{1,b}(\mathbf{k},\tau)+\frac{2}{\tau}\theta_{1,b}(\mathbf{k},\tau)+\frac{6}{\tau^2}
\delta_1(\mathbf{k},\tau)
\nonumber\\
&&-k^2c_s^2(\tau)\delta_{1,b}(\mathbf{k},\tau)=0,
\label{eq:euler_linear1}
\end{eqnarray} 
where the subscripts ``1'' mean that these quantities denote the
first-order perturbations, and
$\delta_1=f_c\delta_{1,c}+f_b\delta_{1,b}$.
We rewrite eq.~[\ref{eq:euler_linear1}] as
\begin{eqnarray}
& &\dot{\theta}_{1,b}(\mathbf{k},\tau)+\frac{2}{\tau}\theta_{1,b}(\mathbf{k},\tau)
\nonumber\\
& &+\frac{6}{\tau^2}\left[
\delta_1(\mathbf{k},\tau)-\frac{k^2c_s(\tau)^2\tau^2}{6}\delta_{1,b}(\mathbf{k},\tau)
\right]=0,\nonumber\\
& & \dot{\theta}_{1,b}(\mathbf{k},\tau)+\frac{2}{\tau}\theta_{1,b}(\mathbf{k},\tau)
\nonumber\\
& &+\frac{6}{\tau^2}
\left[\delta_1(\mathbf{k},\tau)-\frac{k^2}{k^2_J}\delta_{1,b}(\mathbf{k},\tau)\right]=0,
\label{eq:euler_linear}
\end{eqnarray}
where we have used the usual definition of the Jeans wavenumber, $k_J$:
\begin{equation}
k_J(\tau)\equiv \frac{\sqrt{6}}{c_s(\tau) \tau}.
\end{equation}
The Jeans wavenumber divides the solutions for $\delta_{1,b}$ into two
classes: the growing solution for $k\ll k_J$, and the oscillatory
solution for $k\gg k_J$, {\it when there is no CDM,} i.e., $f_b=1$ and
$\delta_1=\delta_{1,b}$. When $\delta_1\ne \delta_{1,b}$, the Jeans
wavenumber does not provide a dividing scale for 
the solutions of $\delta_{1,b}$. 

The Jeans wavenumber depends on the temperature of the matter component
``b''  as $k_J\propto T_b^{-1/2}\tau^{-1}$; thus, $k_J$ depends on time
in general, $k_J=k_J(\tau)$. However, in order to simplify the problem
and obtain physical insights into the effects of pressure on the
non-linear growth of structure, we shall assume that $k_J$ is
independent of time, which requires that the matter temperature evolve
as if the matter were coupled to radiation, $T_b\propto 1/a\propto
1/\tau^2$. This is not a realistic assumption especially in a low
redshift universe where baryons are decoupled from the radiation
background and neutrinos are non-relativistic - in both cases the temperature
evolves as $T_b\propto 1/a^2\propto 1/\tau^4$ and thus $k_J$ evolves as
$k_J\propto \tau\propto a^{1/2}$, for the adiabatic evolution.

We shall solve the above coupled linear equations iteratively: as 
CDM is always the most dominant source of gravity, the zeroth-order
iterative solution may be found by setting $\delta_1\rightarrow
\delta_{1,c}$ (i.e., $f_c\rightarrow 1$). We find the solution for the
ratio of the density contrasts, which is often called the ``Jeans filtering
function,'' \citep{gnedin/hui:1998}
\begin{equation}
 g_1(\mathbf{k},\tau)\equiv
  \frac{\delta_{1,b}(\mathbf{k},\tau)}{\delta_{1,c}(\mathbf{k},\tau)},
\end{equation}
which should be a decreasing function of $k$ due to the effect of
pressure. At the zeroth-order of iteration, the CDM density contrast
grows as 
\begin{equation}
 \delta_{1,c}^{(0)}(\mathbf{k},\tau)\propto a\propto \tau^2,
\end{equation}
 and thus the
equation for $g_1$ simplifies to
\begin{eqnarray}
\ddot{g}_1^{(0)}(\mathbf{k},\tau)+\frac{6}{\tau}\dot{g}_1^{(0)}(\mathbf{k},\tau)
+\frac{6}{\tau^2}\left(1+\frac{k^2}{k^2_J}\right)g_1^{(0)}(\mathbf{k},\tau)
=\frac{6}{\tau^2}.\nonumber\\
\label{eq:zerothgeq}
\end{eqnarray}
The solution for $g_1(\mathbf{k},\tau)$ must be normalized such that
$g_1(k,\tau)\rightarrow 1$ as $k\rightarrow 0$. We find
\begin{eqnarray}
g_1^{(0)}(k,\tau)=\frac{1}{1+\frac{k^2}{k^2_J}}+\mathcal{O}\left(\tau^{m(k)}\right),
\end{eqnarray}
where 
\begin{equation}
 m(k)\equiv -\frac52\left[1\pm\sqrt{1-\frac{24}{25}\left(1+\frac{k^2}{k^2_J}\right)}\right].
\end{equation}
 The second term is a decaying mode, whose amplitude is set by the
 initial condition, e.g., at the epoch when the baryon temperature was
 raised (by, say, cosmic reionization)  to the point where the pressure
 became important, or at the epoch when the neutrinos became non-relativistic.

Ignoring the decaying mode (although we shall come back to this later), 
we have the
 zeroth-order solution:
\begin{eqnarray}
g_1^{(0)}(k)=\frac{1}{1+\frac{k^2}{k^2_J}}.
\label{eq:g(0)_1}
\end{eqnarray}

At the first-order iteration we have the pressure feedback on the growth
of CDM. The evolution of $\delta_{1,c}^{(1)}$ depends on $k$, and
is given by
\begin{eqnarray}
\delta_{1,c}^{(1)}(k,\tau)\propto \tau^{n(k)},
\end{eqnarray}
where
\begin{eqnarray}
n(k)&\equiv& \frac12\left[-1\pm5\sqrt{1-\frac{24}{25}f_b(1-g_1^{(0)}(k))}\right]
\nonumber\\
&\simeq&\left\{\begin{array}{rr}2- \frac65 f_b[1-g^{(0)}_1(k)]\\
-3+\frac65 f_b[1-g^{(0)}_1(k)]\end{array}\right..
\end{eqnarray}
The second equality is valid for $f_b[1-g^{(0)}_1(k)]\ll 1$.
The growing mode solution is given by
\begin{eqnarray}
n_+(k)=2-\frac65f_b[1-g^{(0)}_1(k)].
\label{eq:n+}
\end{eqnarray}
As $g^{(0)}(k)\rightarrow 1$ and $0$ for $k\rightarrow 0$ and $\infty$,
respectively, the large-scale and small-scale limits of the growing mode
solution is \citep[see, e.g., Sec.~8.3 of ][for a recent review]{weinberg:COS}
\begin{eqnarray}
\delta_{1,c+}^{(1)}(k,\tau)&\propto&\tau^2\propto a, \ \ \ \ \ \ \ \ \ \
 \ \ \ \ \ \ \ \ \ \ \ k\ll k_J,\\
\delta_{1,c+}^{(1)}(k,\tau)&\propto&\tau^{2-\frac65f_b}\propto
 a^{1-\frac35f_b},\ \ \ \ \ \ \ \ k\gg k_J.
\end{eqnarray}
The growth of $\delta_{1,c}$ on the spatial scales below the Jeans scale is
suppressed relative to that of the large-scale modes. 

Taking the first order iteration solution for $\delta^{(1)}_{1,c+}$ into account,
the first order iteration equation for $g_1^{(1)}$ is 
\begin{eqnarray}
&&\ddot{g}_1^{(1)}(k,\tau)+\frac1{\tau}\left[1+5\sqrt{1-\frac{24}{25}f_b(1-g^{(0)}_1(k))}\right]
\dot{g}_1^{(1)}(k,\tau)
\nonumber\\
&&+\frac{6}{\tau^2}\left[1+\frac{k^2}{k^2_J}-f_b(2-g_1^{(0)}(k))\right]g_1^{(1)}(k,\tau)
 =\frac{6(1-f_b)}{\tau^2},\nonumber\\ 
\end{eqnarray}
whose growing mode solution (with the normalization that
$g_1^{(1)}\rightarrow 1$ for $k\rightarrow 0$) is
\begin{eqnarray}
\nonumber
g^{(1)}_1(k)&=&\frac{1-f_b}{1+\frac{k^2}{k^2_J}-f_b[2-g^{(0)}_1(k)]}\\
&=&\frac{1-f_b}{1-f_b+\frac{k^2}{k^2_J}\left(1-\frac{f_b}{1+k^2/k_J^2}\right)}.
\end{eqnarray}
This iteration converges quickly for $f_b<0.5$, and further iterations are not necessary.
The largest difference between $g^{(0)}_1(k)$ and $g^{(1)}_1(k)$ occurs
as $k/k_J\to\infty$, and is 100\% for $f_b=0.5$. If the component
``b'' is identified with baryons, $f_b\simeq 1/6$, and the
difference is reduced to $\sim 20$\%.
The difference between $g^{(1)}_1(k)$ and $g^{(2)}_1(k)$ occurs at
$k\sim k_J$, and is $\sim4$\% for $f_b=0.5$, and 0.2\% for $f_b\simeq 1/6$.
The difference is much smaller for neutrinos.

To simplify the subsequent analysis, we shall adopt the zeroth-order
iterative solution for the filtering function,
$g_1^{(0)}=1/(1+k^2/k_J^2)$, and the first-order iterative solution 
for the CDM growth factor, eq.~(\ref{eq:n+}),
as the solution at the
first-order in perturbations. This solution is sufficiently accurate for
our obtaining the physical insights. 

\begin{figure}[t]
\centering
\rotatebox{0}{%
  \includegraphics[width=8cm]{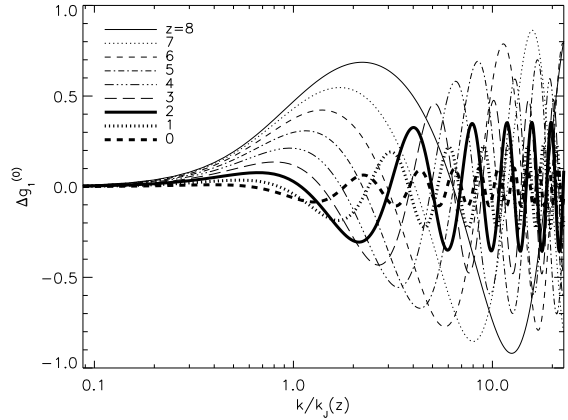}
}%
\caption{%
Decaying mode solution for the linear filtering function at the
 zeroth-order iteration ($f_c\rightarrow 1$), 
$\Delta g_1^{(0)}(k,\tau)\equiv g_1^{(0)}(k,\tau)-1/(1+k^2/k_J^2)$,
where $g_1^{(0)}(k,\tau)$ is the numerical solution of
 eq.~(\ref{eq:zerothgeq}), with the initial conditions given by
$g_1^{(0)}(k,\tau_*)=1$ and $\dot{g}_1^{(0)}(k,\tau_*)=0$
where $\tau_*$ is the conformal time at $z_*=10$. 
The top and bottom lines at $k/k_J\sim 1$ are at $z=8$
 and 0, respectively, and the other lines correspond to the
 intermediate redshifts.
}%
\label{gn_vs_g_diff}
\end{figure}

Let us comment on the decaying mode that we have ignored in obtaining
eq.~[\ref{eq:g(0)_1}]. This decaying mode is an oscillatory function at
$k/k_J> 1/(2\sqrt{6})\simeq 0.2$, representing the acoustic oscillation
of the pressured component \citep{nusser:2000}. 
While this term is a decaying mode, it 
decays slowly, and is not quite negligible even at low redshift.
We show the decaying mode at the zeroth-order iterative solution in
Fig.~\ref{gn_vs_g_diff},  
\begin{equation}
\Delta g_1^{(0)}(k,\tau)\equiv g_1^{(0)}(k,\tau)-\frac1{1+\frac{k^2}{k_J^2}},
\end{equation}
 assuming that the pressure became important at $z_*=10$.
This figure shows that the decaying mode remains important even until
$z\sim 0$; thus, technically speaking, ignoring the decaying mode
results in an inaccurate form of the filtering function.
Nevertheless, we shall ignore it and adopt $g_1(k)=1/(1+k^2/k_J^2)$.

The exact form of $g_1(k,\tau)$ is not so important for our purposes. 
The main goal of this paper is to study how non-linearities affect this
function. In other words, we are interested in how the higher-order
filtering functions, $g_n(\mathbf{k},\tau)$, are related to the linear one,
$g_1(k,\tau)$. One may use any forms of $g_1(k,\tau)$ for a
better accuracy, depending on the problem (baryons or neutrinos).
\subsection{Second and Third Order Solutions}

For the higher order ($n$-th order) density perturbations and
velocity-divergence fields, we  define the
Jeans filtering functions such that
\begin{eqnarray}
&&g_n(\mathbf{k},\tau)\equiv \frac{\delta_{n,b}(\mathbf{k},\tau)}{\delta_{n,c}(\mathbf{k},\tau)},
\\
&&h_n(\mathbf{k},\tau)\equiv \frac{\theta_{n,b}(\mathbf{k},\tau)}{\theta_{n,c}(\mathbf{k},\tau)}.
\label{eq:defn_g_h}
\end{eqnarray}

Assuming that CDM dominates the gravitational potential, we find the
zeroth-order iteration ansatz in an EdS universe:
\begin{eqnarray}
\delta_b(\mathbf{k},\tau)&=&\sum^{\infty}_{n=1}a^n(\tau)\delta_{n,c}(\mathbf{k})g_n(\mathbf{k},\tau),\label{eq:rec_d_b}\\
\theta_b(\mathbf{k},\tau)&=&\sum^{\infty}_{n=1}\dot{a}(\tau)a^{n-1}(\tau)\theta_{n,c}(\mathbf{k})h_n(\mathbf{k},\tau).
\label{eq:rec_t_b}
\end{eqnarray}
Detailed derivations of the non-linear filtering functions at the second
order, $g_2(\mathbf{k},\tau)$, and the third order,
$g_3(\mathbf{k},\tau)$, are given in Appendix
\S~\ref{sec:3pt_baryon_app}. 
The second-order solution is
\begin{equation}
g_2(\mathbf{k},\tau)=\frac{\frac{10}3
-\frac73\left[1-\frac{\delta_{2,c}'(\mathbf{k})}{\delta_{2,c}(\mathbf{k})}
\right]
}
{\frac{10}3+\frac{k^2}{k_J^2}}
+\mathcal{O}(\tau^{-9/2}),
\end{equation}
where 
\begin{eqnarray}
\delta_{2,c}(\mathbf{k})&=&\frac{1}{(2\pi)^3}
\int d\mathbf{q}F_2^{(s)}(\mathbf{q},\mathbf{k}-\mathbf{q})\nonumber\\
& &\times
\delta_{1,c}(\mathbf{q})\delta_{1,c}(\mathbf{k}-\mathbf{q}),\\
\delta'_{2,c}(\mathbf{k})&=&\frac{1}{(2\pi)^3}
\int d\mathbf{q}\left[F_2^{(s)}(\mathbf{q},\mathbf{k}-\mathbf{q})+\frac{3}{14}\frac{k^2}{k_J^2}\right]\nonumber\\
& &\times
g_1(\mathbf{q})g_1(\mathbf{k}-\mathbf{q})
\delta_{1,c}(\mathbf{q})\delta_{1,c}(\mathbf{k}-\mathbf{q}),
\end{eqnarray}
and $F_2^{(s)}$ is a mathematical function given by 
eq.~[\ref{eq:f2s_app}].
The third-order solution is
\begin{equation}
g_3(\mathbf{k},\tau)=\frac{7-6\left[1-\frac{\delta_{3,c}'(\mathbf{k})}{\delta_{3,c}(\mathbf{k})}\right]}
{7+\frac{k^2}{k^2_J}}+\mathcal{O}(\tau^{-13/2}),
\end{equation}
where 
\begin{eqnarray}
\delta_{3,c}(\mathbf{k})&=&
\frac1{(2\pi)^6}\int\!\int\!\int
d\mathbf{q}_1d\mathbf{q}_2d\mathbf{q}_3
\delta_D(\mathbf{k}-\mathbf{q}_1-\mathbf{q}_2-\mathbf{q}_3)\nonumber\\
& &\times
F_3^{(s)}(\mathbf{q}_1,\mathbf{q}_2,\mathbf{q}_3)
\delta_{1,c}(\mathbf{q}_1)\delta_{1,c}(\mathbf{q}_2)
\delta_{1,c}(\mathbf{q}_3),\\
\delta'_{3,c}(\mathbf{k})&=&
\frac1{(2\pi)^6}\int\!\int\!\int
d\mathbf{q}_1d\mathbf{q}_2d\mathbf{q}_3
\delta_D(\mathbf{k}-\mathbf{q}_1-\mathbf{q}_2-\mathbf{q}_3)\nonumber\\
& &\times
{\mathcal F}_3^{(s)}(\mathbf{q}_1,\mathbf{q}_2,\mathbf{q}_3)
\delta_{1,c}(\mathbf{q}_1)\delta_{1,c}(\mathbf{q}_2)
\delta_{1,c}(\mathbf{q}_3),
\end{eqnarray}
and $F_3^{(s)}$ and ${\mathcal F}_3^{(s)}$ are mathematical functions given by 
eqs.~[\ref{eq:f3s_app}] and [\ref{eq:f3prims_app}], respectively.
One may check that these functions are properly normalized, i.e.,
$g_n\rightarrow 1$ as $k\rightarrow 0$, using
$\delta_{2,c}'\rightarrow \delta_{2,c}$ and $\delta_{3,c}'\rightarrow
\delta_{3,c}$ as $k\rightarrow 0$.

Ignoring the decaying modes, let us rewrite $g_2$ and $g_3$ as
\begin{eqnarray}
g_2(\mathbf{k})&=&\frac{1
-\frac7{10}\left[1-\frac{\delta_{2,c}'(\mathbf{k})}{\delta_{2,c}(\mathbf{k})}
\right]
}
{1+\frac3{10}\frac{k^2}{k_J^2}},\\
g_3(\mathbf{k})&=&
\frac{1-\frac{6}{7}\left[1-\frac{\delta_{3,c}'(\mathbf{k})}{\delta_{3,c}(\mathbf{k})}\right]
}
{1+\frac17\frac{k^2}{k^2_J}}.
\end{eqnarray}
These results may be interpreted as, roughly speaking, the non-linear
filtering functions having smaller effective filtering scales 
(larger filtering wavenumbers): 
$k_J\rightarrow \tilde{k}_J=\sqrt{\frac{10}{3}}k_J$ for the second order,
$k_J\to \tilde{k}_J=\sqrt{7}k_J$ for the third order, and
$k_J\to\tilde{k}_J= \sqrt{\frac23n(n+\frac12)}k_J$ for the $n$-th order
perturbations. In other words, the higher-order solutions for
$\delta_{n,b}$ are less suppressed relative to the CDM solutions.
In the next section we shall quantify this effect in more detail.
\section{Power Spectrum}
\label{sec:power_baryon}
In this section, we calculate the non-linear matter power spectrum using
the results obtained in the previous section.
The total matter fluctuation, $\delta$, is given by 
$\delta=f_c\delta_c+f_b\delta_b$, and thus the total matter power
spectrum, $P_{tot}(k)$, is given by the sum of three contributions:
\begin{equation}
 P_{tot}(k,\tau) = f_c^2 P_c(k,\tau) + f_cf_bP_{bc}(k,\tau) + f_b^2P_b(k,\tau),
\end{equation}
where $P_c(k)$ and $P_b(k)$ are the power spectra of the CDM and another
matter component with pressure, respectively, and $P_{bc}(k)$ is the
cross-correlation power spectrum. Each term is the sum of the linear
part, $P_{11}(k,\tau)$, and the non-linear parts, $P_{22}(k,\tau)$ and
$P_{13}(k,\tau)$: 
\begin{equation}
 P_i(k,\tau) = P_{11,i}(k,\tau) + P_{22,i}(k,\tau) + 2P_{13,i}(k,\tau),
\end{equation}
where $i=(c,b,bc)$.

The 3PT power spectrum of CDM has been found in the literature
\citep[see ][for a review]{bernardeau/etal:2002}:
\begin{eqnarray}
P_{22,c}(k,\tau) &=&
2\int \frac{d\mathbf{q}}{(2\pi)^3}
P_{11,c}(q,\tau)P_{11,c}(\left|\mathbf{k}-\mathbf{q}\right|,\tau)
 \nonumber\\
& &\times\left[F^{(s)}_2(\mathbf{q},\mathbf{k}-\mathbf{q})\right]^2,
\end{eqnarray}
where $F^{(s)}_2$ is a mathematical function given by
eq.~[\ref{eq:f2s_app}], and 
\begin{eqnarray}
P_{13,c}(k,\tau)&=&
\frac{2\pi}{252}k^2P_{11,c}(k,\tau)\int_0^{\infty}\frac{dq}{(2\pi)^3}
P_{11,c}(q,\tau)\nonumber\\
& \times&\left[
50\frac{q^2}{k^2}
-21\frac{q^4}{k^4}-79+6\frac{k^2}{q^2}
\right.\nonumber\\
& &\left.+\frac32
\frac{(q^2-k^2)^3(2k^2+7q^2)}{k^5q^3}\ln{\frac{k+q}{\left|k-q\right|}}
\right].
\end{eqnarray}
See Appendix \S~\ref{sec:3pt_cdm_app} for the detailed derivations.

Here, we have implicitly generalized the results from an EdS universe to
general cosmological models, by writing 
\begin{eqnarray}
&&\frac{a^2(\tau)}{a^2(\tau_i)} P_{11}(k,\tau_i)
\to
\nonumber\\
&&P_{11}(k,\tau)\!
=\! \frac{D^2(\tau)}{D^2(\tau_i)}\! 
\left(\frac{\delta_{1,c+}^{(1)}(k,\tau)/\delta_{1,c+}^{(0)}(k,\tau)}
{\delta_{1,c+}^{(1)}(k,\tau_*)/\delta_{1,c+}^{(0)}(k,\tau_*)}\right)^2
\!\!P_{11}(k,\tau_i),
\label{eq:generalization}
\nonumber\\
\end{eqnarray}
where $\tau_i$ is some arbitrary epoch, $\tau_*$ is the epoch
where the pressure effect becomes non-negligible (i.e., reionization
epoch for baryons and the relativistic to non-relativistic transition epoch
for massive neutrinos), and $D(\tau)$ is the linear
growth factor appropriate to a given cosmological model.
This simple generalization has been shown to provide an excellent
approximation to the full calculation:
see \cite{bernardeau/etal:2002} for models with non-zero curvature
and/or a cosmological constant, and 
\cite{takahashi:2008} for  dynamical dark energy models with a constant
equation of state of dark energy.

The linear spectra of the other contributions, 
 $P_{11,bc}$ and $P_{11,b}$, are given by
\begin{eqnarray}
 P_{11,bc}(k,\tau) &=& g_1(k)P_{11,c}(k,\tau),
\label{eq:p11bc}\\
 P_{11,b}(k,\tau) &=& g_1^2(k)P_{11,c}(k,\tau).
\end{eqnarray}
The non-linear terms, the main results of this paper, are given by
\begin{eqnarray}
 P_{22,bc}(k,\tau)
&=&\frac1{\frac{10}3+\frac{k^2}{k^2_J}}
\left[P_{22,c}(k,\tau)\right.\nonumber\\
&+&\frac{14}{3}
\int \frac{d\mathbf{q}}{(2\pi)^3}
P_{11,c}(q,\tau)P_{11,c}(\left|\mathbf{k}-\mathbf{q}\right|,\tau)
 \nonumber\\
& &\left.\times
F^{(s)}_2(\mathbf{q},\mathbf{k}-\mathbf{q})
{\mathcal F}^{(s)}_2(\mathbf{q},\mathbf{k}-\mathbf{q})
\right],
\label{eq:p22bc}
\end{eqnarray}
\begin{eqnarray}
 P_{22,b}(k,\tau)
&=&\frac1{\left(\frac{10}3+\frac{k^2}{k^2_J}\right)^2}
\left[P_{22,c}(k,\tau)\right.\nonumber\\
&+&\frac{28}{3}
\int \frac{d\mathbf{q}}{(2\pi)^3}
P_{11,c}(q,\tau)P_{11,c}(\left|\mathbf{k}-\mathbf{q}\right|,\tau)
 \nonumber\\
& &\times
F^{(s)}_2(\mathbf{q},\mathbf{k}-\mathbf{q})
{\mathcal F}^{(s)}_2(\mathbf{q},\mathbf{k}-\mathbf{q})\nonumber\\
&+&\frac{98}{9}
\int \frac{d\mathbf{q}}{(2\pi)^3}
P_{11,c}(q,\tau)P_{11,c}(\left|\mathbf{k}-\mathbf{q}\right|,\tau)
 \nonumber\\
& &\left.\times
\left({\mathcal F}^{(s)}_2(\mathbf{q},\mathbf{k}-\mathbf{q})\right)^2\right],
\label{eq:p22b}
\end{eqnarray}
\begin{eqnarray}
 P_{13,bc}(k,\tau)
&=&\frac12\left[\left(g_1(k)
+\frac1{7+\frac{k^2}{k^2_J}}\right)P_{13,c}(k,\tau)\right.\nonumber\\
&+&\frac{18}{7+\frac{k^2}{k^2_J}}P_{11,c}(k,\tau)
\int\frac{d\mathbf{q}}{(2\pi)^3}
{\mathcal F}^{(s)}_3(\mathbf{q},-\mathbf{q},\mathbf{k})\nonumber\\
& &\left.\times
P_{11,c}(q,\tau)\right],
\label{eq:p13bc}
\end{eqnarray}
\begin{eqnarray}
 P_{13,b}(k,\tau)
&=& \frac{g_1(k)}{7+\frac{k^2}{k^2_J}}\left[P_{13,c}(k,\tau)+18P_{11,c}(k,\tau)\right.\nonumber\\
& &\left.\times
\int\frac{d\mathbf{q}}{(2\pi)^3}{\mathcal F}^{(s)}_3(\mathbf{q},-\mathbf{q},\mathbf{k})P_{11,c}(q,\tau)
\right].
\label{eq:p13b}
\end{eqnarray}
See Appendix \S~\ref{sec:3pt_power_app} for the detailed derivations.

\begin{figure}[t]
\centering
\rotatebox{0}{%
  \includegraphics[width=8cm]{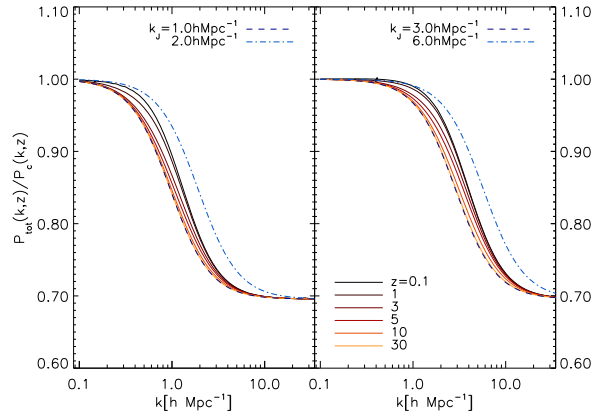}
}%
\caption{%
Ratio of the total matter power spectrum, $P_{tot}(k,z)$, to the CDM
 part, $P_c(k,z)$, at $z=0.1$ (top), 1, 3, 5, 10, and 30 (bottom).
(Left) The input Jeans wavenumber of $k_J=1~h~{\rm Mpc}^{-1}$.
(Right) $k_J=3~h~{\rm Mpc}^{-1}$. The dashed lines show the ratios
 calculated from the linear theory, whereas the dot-dashed lines show
 the linear calculations with $k_J=2$ and $6~h~{\rm Mpc}^{-1}$ for the
 left and right panels respectively, to show that the actual filtering wavenumbers,
 predicted by the 3PT calculations, can
 be $\sim$40\% as large as the linear filtering wavenumber at low redshift.
}%
\label{3pt_baryon_ov_cdm}
\end{figure}

How would $P_{tot}(k)$ compare with the CDM part, $P_{c}(k)$? 
\begin{itemize}
 \item In the linear limit, we should recover
$P_{tot}(k)/P_{c}(k)\rightarrow [f_c+f_bg_1(k)]^2$, which approaches unity as
$k\rightarrow 0$.
 \item  In the very small scale limit ($k\rightarrow \infty$),
the pressured component is completely smooth ($\delta_{b}(k)\rightarrow
	0$) because $g_1(k)\rightarrow 0$; thus, 
$P_{tot}(k)/P_{c}(k)$ approaches a constant value, $f_c^2$. 
 \item In the intermediate regime, especially at  the transition
scale between the super-Jeans scale ($k<k_J$) and the sub-Jeans scale
($k>k_J$), the shape of $P_{tot}(k)/P_{c}(k)$ is significantly
distorted away from the linear prediction. Non-linear clustering of the
pressured component adds power at $k\sim k_J$, which shifts the
effective filtering scale to smaller spatial scales as we go to lower redshifts.
\end{itemize}
In Fig.~\ref{3pt_baryon_ov_cdm} we show the ratio, 
$P_{tot}(k,z)/P_{c}(k,z)$ (solid lines), for different redshifts
($z=0.1$, 1, 3, 5, 10, and 30), and different $k_J$ ($k_J=1$ and
$3~h~{\rm Mpc}^{-1}$ for the left and right panels, 
respectively). 
In the linear regime (see the bottom
lines, $z=30$) the ratio agrees with the linear prediction shown by the
dashed lines. As we go to lower redshifts, we find that the filtering 
wavenumbers continue to shift to larger values, i.e., the filtering
scales continue to shift to smaller spatial scales as we go to lower redshifts.
This effect cannot be predicted from the linear theory, where
all the modes evolve in the same way.
\section{Comparison with Approximate Treatment of Saito et al. (2008)}
\label{sec:semi-analytic}
The non-linear power spectrum with a significant contribution from a
pressured component has not been studied very much in the literature,
with one exception. \citet{saito/etal:2008} (hereafter, STT) have
studied effects of 
massive neutrinos on the non-linear matter power spectrum using 3PT
\citep[also see][]{wong:2008,lesgourgues/etal:2009}.
However, their treatment is not satisfactory: they have entirely ignored
non-linearities in neutrinos, but approximated the neutrino
perturbations as linear perturbations. More precisely, they calculated
the non-linear matter power spectrum as 
\begin{equation}
P^{\rm
 STT}_{tot}(k,z)=f_c^2P_c(k,z)+2f_cf_{\nu}P_{11,\nu c}(k,z)+f_{\nu}^2P_{11,\nu}(k,z).
\end{equation}
In our language this leads to 
 \begin{equation}
P^{\rm
 STT}_{tot}(k,z)=
f_c^2P_c(k)+[2f_cf_{\nu}g_1(k)+f_{\nu}^2g_1^2(k)]P_{11,c}(k,z).
\label{semi_analytic_power}
\end{equation}
Here, we have replaced the subscripts ``b'' with ``$\nu$'' to avoid
confusion in notation.

\begin{figure}[t]
\centering
\rotatebox{0}{%
  \includegraphics[width=8cm]{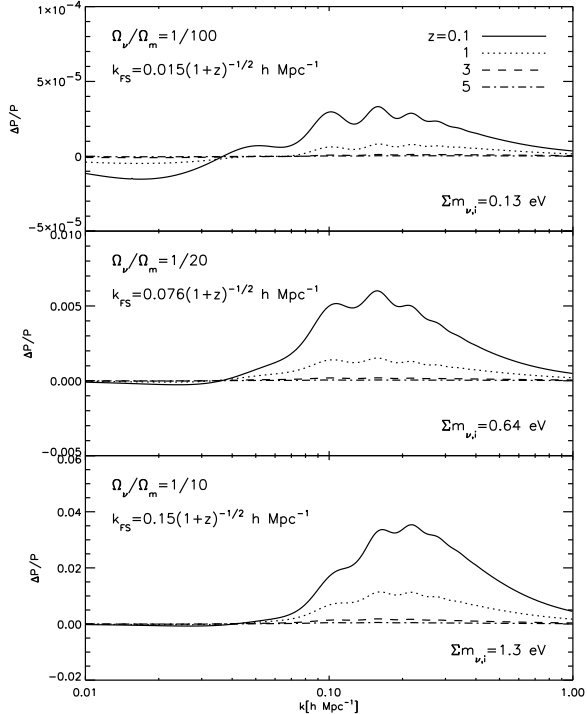}
}%
\caption{%
Fractional difference between our full calculation and the approximation
 used by \cite{saito/etal:2008} (STT), $[P_{tot}(k)-P^{\rm
 STT}_{tot}(k)]/P_{tot}(k)$, for $\Omega_\nu/\Omega_m=1/100$ (top), 
$1/20$ (middle), and $1/10$ (bottom), which
 corresponds to $\sum m_\nu\simeq 0.13$, $0.64$, and $1.3$~eV, respectively.
}%
\label{3pt_baryon_vs_semi_frac_diff}
\end{figure}

How accurate is the STT approximation? 
To study this, we compare eq.~[\ref{semi_analytic_power}] to the full
calculation given in the previous section.
Figure \ref{3pt_baryon_vs_semi_frac_diff} shows the fractional
difference between our full calculation and STT's approximation,
$[P_{tot}(k)-P^{\rm STT}_{tot}(k)]/P_{tot}(k)$, for 
$\Omega_\nu/\Omega_m=1/10$, $1/20$, and $1/100$, which correspond to the
sum of neutrino masses of $\sum_i m_{\nu,i}\simeq 1.3$, $0.64$, and $0.13$~eV,
respectively, where $i=(e,\mu,\tau)$.
We find that 
STT's approximation clearly underestimates the power at $k\approx
k_{FS}$, where $k_{FS}$ is the neutrino free-streaming scale, or it is
the Jeans wavenumber computed with the velocity dispersion of the
neutrinos. More precisely,
\begin{equation}
 k_{FS,i}(\tau) \equiv \frac{\sqrt{6}}{\sigma_{\nu,i}(\tau)\tau},
\end{equation}
in an EdS universe, 
where $\sigma_{\nu,i}^2(\tau)$ is the velocity dispersion of neutrino
species $i$ \citep[see, e.g., Appendix A.3 of ][]{takada/komatsu/futamase:2006}.

\begin{figure}[t]
\centering
\rotatebox{0}{%
  \includegraphics[width=8cm]{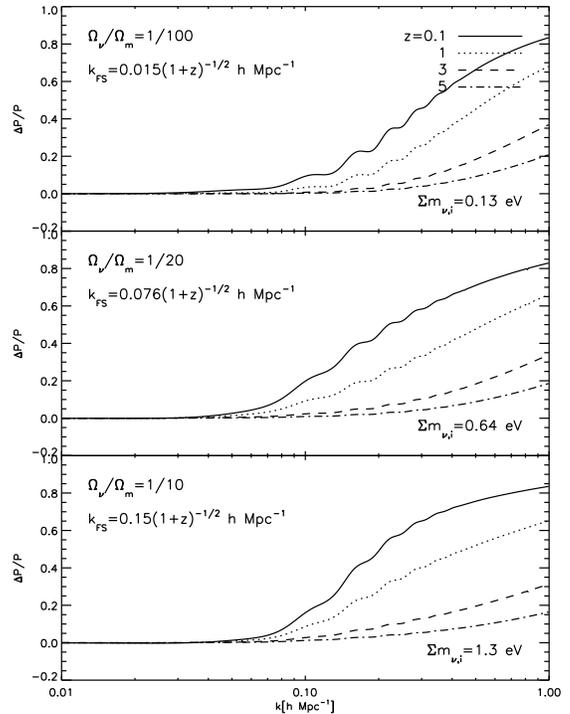}
}%
\caption{%
Fractional difference between the non-linear neutrino power spectrum,
$P_\nu(k)$, and the linear power spectrum, $P_\nu^{lin}(k)$, 
 $[P_{\nu}(k)-P^{\rm
 lin}_{\nu}(k)]/P_{\nu}(k)$, for $\Omega_\nu/\Omega_m=1/100$ (top), 
$1/20$ (middle), and $1/10$ (bottom), which
 corresponds to $\sum m_\nu\simeq 0.13$, $0.64$, and $1.3$~eV, respectively.
}%
\label{3pt_vs_semi_frac_diff_neutrino_mod}
\end{figure}

One may argue that STT's approximation should be better for a smaller
neutrino mass: the errors in the total matter power spectrum are 3.5\%,
0.6\%, and 0.003\%  
$\sum_i m_{\nu,i}=1.3$, $0.64$, and $0.13$~eV, respectively, at $z=0.1$;
however,  
our results indicate that their approximation is conceptually
not correct: neutrinos should not be treated as linear perturbations, as
the neutrino velocity dispersion has no effect in suppressing the neutrino
perturbations at and above the free-streaming scale. In other words, the
errors may happen to be small in the {\it total} matter power spectrum for small
neutrino masses because neutrinos contribute only a tiny fraction of the
total matter density anyway,  
but the errors in the neutrino power spectrum are
large. Figure~\ref{3pt_vs_semi_frac_diff_neutrino_mod} shows the
fractional difference between the non-linear neutrino power spectrum,
$P_\nu(k)$, and the linear power spectrum, $P_\nu^{lin}(k)$, i.e., 
$\Delta P/P= [P_\nu(k)-P_\nu^{lin}(k)]/P_\nu(k)$. It is clear that
neutrinos are significantly non-linear, even well below the
free-streaming scale, $k\gg k_{FS}$. Nevertheless, the STT approximation
may still provide a convenient phenomenological tool for calculating the
non-linear 
total matter power spectrum in the presence of massive neutrinos.
\section{Discussions and Conclusions}
\label{sec:discussion}
In this paper, we have obtained the second- and third-order solutions
for the density perturbations in a system consisting of two matter
components with and without the pressure gradient. This is the first
self-consistent analytical calculation, with non-linearities in the pressured
component fully retained up to the 3rd order in perturbations.

As our study is focused on understanding the physics of the
non-linear pressure effect on the matter power spectrum, we have studied
a toy model in which the Jeans wavenumber, $k_J$, is independent of
time. This is equivalent to the temperature of the pressured component
following that of radiation, i.e., $T\propto 1/a$.

Nevertheless, we have found several results that have qualitative
implications for 
the practical applications. We have found that non-linearities in the
pressured component shift the  filtering scale from the well-known
linear filtering scale \citep{gnedin/hui:1998} to a smaller spatial scale
(larger wavenumber) by a
factor depending on the redshift and the Jeans scale. 
In other words, the actual filtering scale for a given sound speed (or
temperature) is {\it smaller} than the linear scale.
Therefore, if one used the linear filtering scale to interpret the
fall-off of, e.g., the flux power spectrum of the Lyman-$\alpha$ forests
\citep{zaldarriaga/etal:2001}, one would underestimate the temperature
of the pressured component. 

How important is this effect? For example, when the Jeans wavenumber is
$k_J=10~h~{\rm Mpc^{-1}}$, our calculation predicts that 
the effective filtering wavenumber is
$\simeq 10$, 12, 13, 13, and $14~h~{\rm Mpc^{-1}}$
at $z=30$, 10, 5, 3, and 1, respectively. While we do not expect
3PT to be valid at such high wavenumbers, our results clearly indicate
that the expected changes in the filtering scale cannot be ignored.
Table \ref{table:1} summarizes the ratios of the effective (actual) and
the linear filtering wavenumbers. Note that the linear filtering wavenumber
is the same as the Jeans wavenumber in our model; thus, we show $k_{F,eff}/k_J$ in
Table \ref{table:1}. We extracted the effective filtering wavenumber,
$k_{F,eff}$, by 
fitting $[f_c+f_b/(1+k^2/k_{F,eff}^2)]^2$ to $P_{tot}(k,z)/P_{c}(k,z)$.
We find that a factor of $1.4$ change in the filtering scale is quite
common over a wide range of redshifts and $k_J$.

A factor of $1.4$ change in the filtering scale changes the inferred
temperature by a factor of two; thus, one implication of our result is
that the temperature of the Inter-galactic Medium (IGM) obtained from
the Lyman-$\alpha$ forests at $z=3$ by \cite{zaldarriaga/etal:2001}
might have been underestimated by a factor of two.

A factor of $1.4$ change in the filtering scale gives a factor of $\sim 3$
change in the filtering mass. Our calculation shows that the actual
filtering mass is similar to the linear one only in high redshifts,
while the former is significantly smaller than the latter in low redshift.
This result is qualitatively similar to those found in
\citet{okamoto/etal:2008} and \citet{hoeft/etal:2006}; however, a quantitative
comparison is not possible, as our results apply only to the system with
a constant Jeans wavenumber.

What is next? As for baryons, we need to extend our formalism for incorporating
a realistic thermal history of the universe with a proper time
dependence of $k_J$. As for neutrinos, we need to incorporate not only
the pressure gradient but also the anisotropic stress in the Euler
equation. To do this we need to solve the Boltzmann equation. 
Nevertheless, our results presented in this paper show that
neutrinos are significantly non-linear, even well below 
the free-streaming scale.

\begin{deluxetable}{ccccccc}
\tablecaption{Ratio of the effective and the linear filtering scales, 
$k_{F,eff}/k_J$}
\tablehead{\colhead{$k_J$} & \colhead{z=0.1} & \colhead{1.0} & \colhead{3.0} & \colhead{5.0} & \colhead{10} & \colhead{30} \\ 
\colhead{($h~{\rm Mpc^{-1}}$)} & \colhead{} & \colhead{} & \colhead{} & \colhead{} & \colhead{} & \colhead{} } 
\startdata
0.1 & 1.08 & 1.04 & 1.01 & 1.00 & 1.00 & 1.00 \\
0.5 & 1.37 & 1.21 & 1.07 & 1.03 & 1.01 & 1.00 \\
1.0 & 1.43 & 1.32 & 1.14 & 1.08 & 1.03 & 1.00 \\
3.0 & 1.41 & 1.38 & 1.28 & 1.20 & 1.08 & 1.01 \\
5.0 & 1.40 & 1.39 & 1.32 & 1.24 & 1.12 & 1.02 \\
10 & 1.41 & 1.40 & 1.35 & 1.29 & 1.16 & 1.03
\enddata
\tablecomments{%
This table shows the ratios of the effective ($k_{F,eff}$) and the linear ($k_J$)
 filtering 
 scales for different redshifts and $k_J$.
The ratios are closer to unity at higher redshifts because non-linearities
 are  weaker.
}%
\label{table:1}
\end{deluxetable}
This material is based in part upon work 
supported by the Texas Advanced Research Program under 
Grant No. 003658-0005-2006, by NASA grants NNX08AM29G 
and NNX08AL43G, and by NSF grant AST-0807649.
E.~K. acknowledges support from an Alfred P. Sloan Research Fellowship. 
\appendix
\section{3PT for CDM}
\label{sec:3pt_cdm_app}

The continuity, Euler, and Poisson equations of CDM 
are given by

\begin{itemize}
\item Continuity equation:
\begin{equation}
\dot{\delta}(\mathbf{x},\tau)+\mathbf{\nabla}\cdot[(1+\delta(\mathbf{x},\tau))\mathbf{v}(\mathbf{x},\tau)]=0,
\label{eq:continuity}
\end{equation}

\item Euler equations:
\begin{equation}
\dot{\mathbf{v}}(\mathbf{x},\tau)+[\mathbf{v}(\mathbf{x},\tau)\cdot\mathbf{\nabla}]\mathbf{v}(\mathbf{x},\tau)
=-\frac{\dot{a}}{a}\mathbf{v}(\mathbf{x},\tau)-\mathbf{\nabla}\phi(\mathbf{x},\tau),
\label{eq:euler}
\end{equation}

\item Poisson equation (for an EdS universe):
\begin{equation}
\nabla^2\phi(\mathbf{x},\tau)=\frac{6}{\tau^2}\delta(\mathbf{x},\tau).
\label{eq:poisson}
\end{equation}
\end{itemize}

First, we take the divergence of Eq.[\ref{eq:euler}] 
and substitute Eq.[\ref{eq:poisson}]. Moving all the non-linear terms to the
RHS of the equations, we find
\begin{eqnarray}
&&\dot{\delta}(\mathbf{x},\tau)+\mathbf{\nabla}\cdot\mathbf{v}(\mathbf{x},\tau)
=-\mathbf\nabla\cdot[\delta(\mathbf{x},\tau)\mathbf{v}(\mathbf{x},\tau)],
\label{eq:nl_continuity_app}\\
&&\frac{\partial}{\partial \tau}[\mathbf{\nabla}\cdot\mathbf{v}(\mathbf{x},\tau)]
+\frac{\dot{a}}{a}[\mathbf{\nabla}\cdot\mathbf{v}(\mathbf{x},\tau)]+\frac{6}{\tau^2}\delta(\mathbf{x},\tau)
=-\mathbf{\nabla}\cdot\lbrace[\mathbf{v}(\mathbf{x},\tau)\cdot\mathbf{\nabla}]\mathbf{v}(\mathbf{x},\tau)\rbrace.
\label{eq:nl_euler_app}
\end{eqnarray}
Let us take the Fourier transform of Eqs.~[\ref{eq:nl_continuity_app}] and
[\ref{eq:nl_euler_app}]: 
\begin{eqnarray}
\dot{\tilde{\delta}}(\mathbf{k},\tau)+\tilde{\theta}(\mathbf{k},\tau)
&=&-\frac1{(2\pi)^3}\int\!\int d\mathbf{q}_1d\mathbf{q}_2
\delta_D(\mathbf{q}_1+\mathbf{q}_2-\mathbf{k})
\frac{\mathbf{k}\cdot\mathbf{q}_1}{q_1^2}
\tilde{\theta}(\mathbf{q}_1,\tau)\tilde{\delta}(\mathbf{q}_2,\tau),
\label{eq:f_nl_continuity_app}
\end{eqnarray}
\begin{eqnarray}
\dot{\tilde{\theta}}(\mathbf{k},\tau)+\frac{\dot{a}}{a}\tilde{\theta}(\mathbf{k},\tau)+\frac{6}{\tau^2}
\tilde{\delta}(\mathbf{k},\tau)
&=&-\frac1{(2\pi)^3}\int\int d\mathbf{q}_1d\mathbf{q}_2
\delta_D(\mathbf{q}_1+\mathbf{q}_2-\mathbf{k})
\frac{k^2(\mathbf{q}_1\cdot\mathbf{q}_2)}{2q_1^2q_2^2}
\tilde{\theta}(\mathbf{q}_1,\tau)\tilde{\theta}(\mathbf{q}_2,\tau),
\label{eq:f_nl_euler_app}
\end{eqnarray}
where we have defined $\theta\equiv\mathbf{\nabla}\cdot\mathbf{v}$,
and its Fourier transform is given by
\begin{equation}
\tilde{\mathbf{v}}(\mathbf{k},\tau)=-i\frac{\mathbf{k}}{k^2}\tilde{\theta}(\mathbf{k},\tau).
\end{equation}

One can decompose the solutions of
the non-linear continuity and Euler equations,
$\tilde{\delta}$ and $\tilde{\theta}$, into the sum
of infinite series of $n$-th order perturbations of density and
velocity divergence fields:
\begin{eqnarray}
\tilde{\delta}(\mathbf{k},\tau)&=&\sum^{\infty}_{n=1}a^n(\tau)\delta_n(\mathbf{k}),
\\
\tilde{\theta}(\mathbf{k},\tau)&=&\sum^{\infty}_{n=1}\dot{a}(\tau)a^{n-1}(\tau)\theta_n(\mathbf{k}),
\label{eq:infinite_sum_cdm_app}
\end{eqnarray}
respectively. Note that, strictly speaking, this particular decomposition,
a decomposition into a series with powers of $a(\tau)$, 
is valid only for an EdS universe. However, generalization to arbitrary
cosmological models can be done in the end by replacing $a(\tau)$ with
the appropriate growth factor, $D(\tau)$
\citep{bernardeau/etal:2002,takahashi:2008}. 

Now, let us solve Eqs.~[\ref{eq:f_nl_continuity_app}] and
[\ref{eq:f_nl_euler_app}] at each order of perturbations.
The $n$-th ($n>1$) term of the Eq.~[\ref{eq:f_nl_continuity_app}] is given by
\begin{eqnarray}
\dot{a}(\tau)a^{n-1}(\tau)
[n\delta_n(\mathbf{k})+\theta_n(\mathbf{k})]
=-\frac1{(2\pi)^3}\int\!\int d\mathbf{q}_1d\mathbf{q}_2
\delta_D(\mathbf{q}_1+\mathbf{q}_2-\mathbf{k})
\frac{\mathbf{k}\cdot\mathbf{q}_1}{q_1^2}
\sum^{n-1}_{m=1}\dot{a}(\tau)a^{n-1}(\tau)
\theta_m(\mathbf{q}_1)\delta_{n-m}(\mathbf{q}_2).
\end{eqnarray}
Dividing both sides by $\dot{a}(\tau)a^{n-1}(\tau)$, one obtains
\begin{eqnarray}
n\delta_n(\mathbf{k})+\theta_n(\mathbf{k})=A_n(\mathbf{k})\label{eq:recursion_a_app},
\end{eqnarray}
where
\begin{eqnarray}
A_n(\mathbf{k})=-\frac1{(2\pi)^3}\int\!\int d\mathbf{q}_1d\mathbf{q}_2
\delta_D(\mathbf{q}_1+\mathbf{q}_2-\mathbf{k})
\frac{\mathbf{k}\cdot\mathbf{q}_1}{q_1^2}
\sum^{n-1}_{m=1}\theta_m(\mathbf{q}_1)\delta_{n-m}(\mathbf{q}_2).
\end{eqnarray}
Similarly, from the Euler equation, Eq.~[\ref{eq:f_nl_euler_app}], one obtains
\begin{eqnarray}
3\delta_n(\mathbf{k})+(1+2n)\theta_n(\mathbf{k})=B_n(\mathbf{k})\label{eq:recursion_b_app},
\end{eqnarray}
where
\begin{eqnarray}
B_n(\mathbf{k})=-\frac1{(2\pi)^3}\int\!\int d\mathbf{q}_1d\mathbf{q}_2
\delta_D(\mathbf{q}_1+\mathbf{q}_2-\mathbf{k})
\frac{k^2(\mathbf{q}_1\cdot\mathbf{q}_2)}{q_1^2q_2^2}
\sum^{n-1}_{m=1}\theta_m(\mathbf{q}_1)\theta_{n-m}(\mathbf{q}_2).
\end{eqnarray}

The forms of Eqs.~[\ref{eq:recursion_a_app}] and
[\ref{eq:recursion_b_app}] indicate that the $n$-th order solutions are
written in terms of the sum of 1-st to $(n-1)$-th order solutions,
with $\delta_1(\mathbf{k})=-\theta_1(\mathbf{k})$.
By solving Eqs.~[\ref{eq:recursion_a_app}] and [\ref{eq:recursion_b_app}]
for $\delta_n$ and $\theta_n$, one obtains
\begin{eqnarray}
&&\delta_n(\mathbf{k})=\frac{(1+2n)A_n(\mathbf{k})-B_n(\mathbf{k})}{(2n+3)(n-1)},
\label{eq:delta_n_app}
\\
&&\theta_n(\mathbf{k})=\frac{-3A_n(\mathbf{k})+nB_n(\mathbf{k})}{(2n+3)(n-1)},
\label{eq:theta_n_app}
\end{eqnarray}
which can be rewritten as
\begin{eqnarray}
\delta_n(\mathbf{k})=\frac{1}{(2\pi)^{3n-3}}
&&\int d\mathbf{q}_1...d\mathbf{q}_n
\delta_D(\mathbf{q}_1+...+\mathbf{q}_n-\mathbf{k})
F_n(\mathbf{q}_1,...,\mathbf{q}_n)
\delta_1(\mathbf{q}_1)...\delta_1(\mathbf{q}_n)\label{eq:recursion_d_app},
\\
\theta_n(\mathbf{k})=-\frac{1}{(2\pi)^{3n-3}}
&&\int d\mathbf{q}_1...d\mathbf{q}_n
\delta_D(\mathbf{q}_1+...+\mathbf{q}_n-\mathbf{k})
G_n(\mathbf{q}_1,...,\mathbf{q}_n)
\delta_1(\mathbf{q}_1)...\delta_1(\mathbf{q}_n)\label{eq:recursion_t_app}.
\end{eqnarray}
Here, the newly defined kernels,  $F_n$ and $G_n$, can be found
from the following recursion relations:
\begin{eqnarray}
F_n(\mathbf{q}_1,...,\mathbf{q}_n)=\sum^{n-1}_{m=1}
\frac{G_m(\mathbf{q}_1,...,\mathbf{q}_m)}{(2n+3)(n-1)}
\left[(1+2n)\frac{\mathbf{k}\cdot\mathbf{q}_1}{q_1^2}
F_{n-m}(\mathbf{q}_{m+1},...,\mathbf{q}_n)
\frac{k^2(\mathbf{q}_1\cdot\mathbf{q}_2)}{q_1^2q_2^2}
G_{n-m}(\mathbf{q}_{m+1},...,\mathbf{q}_n)
\right]\label{eq:fn_app},
\end{eqnarray}
and
\begin{eqnarray}
G_n(\mathbf{q}_1,...,\mathbf{q}_n)=\sum^{n-1}_{m=1}
\frac{G_m(\mathbf{q}_1,...,\mathbf{q}_m)}{(2n+3)(n-1)}
\left[3\frac{\mathbf{k}\cdot\mathbf{q}_1}{q_1^2}
F_{n-m}(\mathbf{q}_{m+1},...,\mathbf{q}_n)
+n\frac{k^2(\mathbf{q}_1\cdot\mathbf{q}_2)}{q_1^2q_2^2}
G_{n-m}(\mathbf{q}_{m+1},...,\mathbf{q}_n)
\right]\label{eq:gn_app},
\end{eqnarray}
with the boundary conditions of $F_1=1=G_1$.
The 2nd-order solutions are 
\begin{eqnarray}
F_2(\mathbf{q}_1,\mathbf{q}_2)&=&\frac57\frac{\mathbf{k}\cdot\mathbf{q}_1}{q^2_1}
+\frac{k^2(\mathbf{q}_1\cdot\mathbf{q}_2)}{7q^2_1q^2_2},
\label{eq:f2_app}\\
G_2(\mathbf{q}_1,\mathbf{q}_2)&=&\frac37\frac{\mathbf{k}\cdot\mathbf{q}_1}{q^2_1}
+\frac{2k^2(\mathbf{q}_1\cdot\mathbf{q}_2)}{7q^2_1q^2_2},
\label{eq:g2_app}
\end{eqnarray}
where $\mathbf{k}=\mathbf{q}_1+\mathbf{q}_2$.
The 3rd-order solutions are
\begin{eqnarray}
F_3(\mathbf{q}_1,\mathbf{q}_2,\mathbf{q}_3)
&=&
\frac{1}{18}
\left[\frac{7\mathbf{k}\cdot\mathbf{q}_1}{q^2_1}F_2(\mathbf{q}_2,\mathbf{q}_3)
+\frac{k^2(\mathbf{q}_1\cdot\mathbf{q}_{23})}{q^2_1q^2_{23}}G_2(\mathbf{q}_2,\mathbf{q}_3)
\right]
+\frac{G_2(\mathbf{q}_1,\mathbf{q}_2)}{18}
\left[\frac{7\mathbf{k}\cdot\mathbf{q}_{12}}{q^2_{12}}
+\frac{k^2(\mathbf{q}_{12}\cdot\mathbf{q}_3)}{q^2_{12}q^2_3}
\right]\label{eq:f3_app},
\end{eqnarray}
where $\mathbf{q}_{ij}\equiv\mathbf{q}_i+\mathbf{q}_j$ and
$\mathbf{k}=\sum\mathbf{q}_i$.

It is often convenient to have the symmetrized forms of the above
kernels.
They are
\begin{eqnarray}
F^{(s)}_2(\mathbf{q}_1,\mathbf{q}_2)&=&\frac12\left[
F_2(\mathbf{q}_1,\mathbf{q}_2)+F_2(\mathbf{q}_2,\mathbf{q}_1)\right],
\label{eq:f2_sym_app}
\\
G^{(s)}_2(\mathbf{q}_1,\mathbf{q}_2)&=&\frac12\left[
G_2(\mathbf{q}_1,\mathbf{q}_2)+G_2(\mathbf{q}_2,\mathbf{q}_1)\right],
\label{eq:g2_sym_app}
\\
F^{(s)}_3(\mathbf{q}_1,\mathbf{q}_2,\mathbf{q}_3)&=&\frac16\left[
F_3(\mathbf{q}_1,\mathbf{q}_2,\mathbf{q}_3)+F_3(\mathbf{q}_1,\mathbf{q}_3,\mathbf{q}_2)
+F_3(\mathbf{q}_2,\mathbf{q}_1,\mathbf{q}_3)
\right.\nonumber\\
& &+\left.F_3(\mathbf{q}_2,\mathbf{q}_3,\mathbf{q}_1)
+F_3(\mathbf{q}_3,\mathbf{q}_1,\mathbf{q}_2)
+F_3(\mathbf{q}_3,\mathbf{q}_2,\mathbf{q}_1)
\right].
\label{eq:f3_sym_app}
\end{eqnarray}
The explicit forms are
\begin{eqnarray}
F^{(s)}_2(\mathbf{q}_1,\mathbf{q}_2)&=&
\frac57+\frac27\frac{(\mathbf{q}_1\cdot\mathbf{q}_2)^2}{q_1^2q_2^2}
+\frac12\frac{(\mathbf{q}_1\cdot\mathbf{q}_2)(q_1^2+q_2^2)}{q_1^2q_2^2},
\label{eq:f2s_app}\\
G^{(s)}_2(\mathbf{q}_1,\mathbf{q}_2)&=&
\frac37+\frac47\frac{(\mathbf{q}_1\cdot\mathbf{q}_2)^2}{q_1^2q_2^2}
+\frac12\frac{(\mathbf{q}_1\cdot\mathbf{q}_2)(q_1^2+q_2^2)}{q_1^2q_2^2},
\label{eq:g2s_app}\\
F^{(s)}_3(\mathbf{q}_1,\mathbf{q}_2,\mathbf{q}_3)&=&
\frac{7}{54}\mathbf{k}\cdot\left[F^{(s)}_2(\mathbf{q}_2,\mathbf{q}_3)
\frac{\mathbf{q}_1}{q_1^2}
+F^{(s)}_2(\mathbf{q}_1,\mathbf{q}_3)\frac{\mathbf{q}_2}{q_2^2}
+F^{(s)}_2(\mathbf{q}_1,\mathbf{q}_2)\frac{\mathbf{q}_3}{q_3^2}\right]
\nonumber\\
&&+\frac{1}{27}k^2\left[
G^{(s)}_2(\mathbf{q}_2,\mathbf{q}_3)\frac{\mathbf{q}_1\cdot\mathbf{q}_{23}}{q_1^2q_{23}^2}+G^{(s)}_2(\mathbf{q}_1,\mathbf{q}_3)\frac{\mathbf{q}_2\cdot\mathbf{q}_{13}}{q_2^2q_{13}^2}
+G^{(s)}_2(\mathbf{q}_1,\mathbf{q}_2)\frac{\mathbf{q}_3\cdot\mathbf{q}_{12}}{q_3^2q_{12}^2}
\right]
\nonumber
\\
&&+\frac{7}{54}\mathbf{k}\cdot\left[G^{(s)}_2(\mathbf{q}_2,\mathbf{q}_3)
\frac{\mathbf{q}_{23}}{q_{23}^2}
+G^{(s)}_2(\mathbf{q}_1,\mathbf{q}_3)\frac{\mathbf{q}_{13}}{q_{13}^2}
+G^{(s)}_2(\mathbf{q}_1,\mathbf{q}_2)\frac{\mathbf{q}_{12}}{q_{12}^2}\right].
\label{eq:f3s_app}
\end{eqnarray}

In order to calculate the next-to-linear-order density power spectrum,
one needs to use the solutions of the density fluctuations up to the 3rd order:
\begin{eqnarray}
(2\pi)^3P(k,\tau)\delta_D(\mathbf{k}+\mathbf{k}^{\prime})&=&
\langle\tilde{\delta}(\mathbf{k},\tau)\tilde{\delta}(\mathbf{k}^{\prime},\tau)\rangle
\nonumber\\
&=&\left<\left(\sum^{\infty}_{m=1}a^m(\tau)\tilde{\delta}_m(\mathbf{k})\right)
\left(\sum^{\infty}_{l=1}a^l(\tau)\tilde{\delta}_l(\mathbf{k}^{\prime})\right)\right>
\nonumber\\
&\simeq&a^2(\tau)\langle\delta_1(\mathbf{k})\delta_1(\mathbf{k}^{\prime})\rangle
+a^4(\tau)\langle\delta_1(\mathbf{k})\delta_3(\mathbf{k}^{\prime})
+\delta_2(\mathbf{k})\delta_2(\mathbf{k}^{\prime})
+\delta_3(\mathbf{k})\delta_1(\mathbf{k}^{\prime})\rangle,
\end{eqnarray}
which yields 
\begin{equation}
P(k,\tau)=
a^2(\tau)P_{11}(k)+a^4(\tau)\left[P_{22}(k)+2P_{13}(k)\right]+\mathcal{O}(\delta^6).
\end{equation}
Here, we have defined the quantity, $P_{ij}(k)$, given by
\begin{eqnarray}
(2\pi)^3P_{ij}(k)\delta_D(\mathbf{k}+\mathbf{k}^{\prime})
=\langle\delta_i(\mathbf{k})\delta_j(\mathbf{k}^{\prime})\rangle.
\end{eqnarray}
The non-linear corrections, $P_{22}(k)$ and $P_{13}(k)$, are 
\begin{eqnarray}
P_{22}(k)=
2\int \frac{d\mathbf{q}}{(2\pi)^3}
P_{11}(q)P_{11}(\left|\mathbf{k}-\mathbf{q}\right|)
\left[F^{(s)}_2(\mathbf{q},\mathbf{k}-\mathbf{q})\right]^2,
\label{eq:p22_app}
\end{eqnarray}
where 
\begin{eqnarray}
F^{(s)}_2(\mathbf{q},\mathbf{k}-\mathbf{q})
&=&\frac57+\frac{1}{14}\left[
\frac{-10q^4+20kq^3\mu-10k^2q^2\mu^2-7k^2q^2+7k^3q\mu}{q^2(k^2+q^2-2kq\mu)}\right],
\end{eqnarray}
and $\mu \equiv \hat{\mathbf{k}}\cdot\hat{\mathbf{q}}$, and
\begin{eqnarray}
P_{13}(k)
=
3P_{11}(k)
\int \frac{d\mathbf{q}}{(2\pi)^3}F^{(s)}_3(\mathbf{q},-\mathbf{q},\mathbf{k})
P_{11}(q).\label{eq:p13_app}
\end{eqnarray}
Using 
\begin{eqnarray}
\int_{-1}^1d\mu F^{(s)}_3(\mathbf{q},-\mathbf{q},\mathbf{k})=
\frac{1}{756}\left[
50-21\frac{q^2}{k^2}-79\frac{k^2}{q^2}+6\frac{k^4}{q^4}+\frac32
\frac{(q^2-k^2)^3(2k^2+7q^2)}{k^3q^5}\ln{\frac{k+q}{\left|k-q\right|}}
\right],
\nonumber
\end{eqnarray}
one obtains \citep{makino/etal:1992}
\begin{eqnarray}
P_{13}(k)&=&
\frac{2\pi}{252}k^2P_{11}(k)\int_0^{\infty}\frac{dq}{(2\pi)^3}P_{11}(q)
\left[
50\frac{q^2}{k^2}-21\frac{q^4}{k^4}-79+6\frac{k^2}{q^2}+\frac32
\frac{(q^2-k^2)^3(2k^2+7q^2)}{k^5q^3}\ln{\frac{k+q}{\left|k-q\right|}}
\right].
\end{eqnarray}

\section{3PT with Pressure}
\label{sec:3pt_baryon_app}
In this Appendix we shall derive the higher-order filtering
functions. We shall solve
Eqs.~[\ref{eq:f_nl_continuity_c}]--[\ref{eq:f_nl_euler_b}]
perturbatively, up to the 3rd-order in perturbations.
The density contrasts and velocity divergence fields of CDM and the
matter with pressure are all expanded into the infinite sum of $n$-th
order perturbations as

\begin{eqnarray}
&&\tilde{\delta}_c(\mathbf{k},\tau)=\sum^{\infty}_{n=1}a^n(\tau)\delta_{n,c}(\mathbf{k}),
\\
&&\tilde{\theta}_c(\mathbf{k},\tau)=\sum^{\infty}_{n=1}\dot{a}(\tau)a^{n-1}(\tau)\theta_{n,c}(\mathbf{k}),
\end{eqnarray}
\begin{eqnarray}
\tilde{\delta}_b(\mathbf{k},\tau)&=&\sum^{\infty}_{n=1}a^n(\tau)\delta_{n,c}(\mathbf{k})g_n(\mathbf{k},\tau),
\label{eq:rec_d_b_app}\\
\tilde{\theta}_b(\mathbf{k},\tau)&=&\sum^{\infty}_{n=1}\dot{a}(\tau)a^{n-1}(\tau)\theta_{n,c}(\mathbf{k})h_n(\mathbf{k},\tau),
\label{eq:rec_t_b_app}
\end{eqnarray}
where $g_n({\mathbf k},\tau)$ and $h_n({\mathbf k},\tau)$ are the
filtering functions for the density and velocity divergence fields, respectively, at
the $n$-th
order. 

With the above series expansion,
Eqs.~[\ref{eq:f_nl_continuity_b}] and [\ref{eq:f_nl_euler_b}] yield 
\begin{eqnarray}
&&\sum_{n=1}^{\infty}\left[
\left(n\dot{a}(\tau)a^{n-1}(\tau)g_n(\mathbf{k},\tau)+a^n(\tau)\dot{g}_n(\mathbf{k},\tau)
\right)\delta_{n,c}(\mathbf{k})
+\dot{a}(\tau)a^{n-1}(\tau)h_n(\mathbf{k},\tau)\theta_{n,c}(\mathbf{k})\right]
\nonumber\\
&&=-\frac1{(2\pi)^3}\int\!\int d\mathbf{q}_1d\mathbf{q}_2
\delta_D(\mathbf{q}_1+\mathbf{q}_2-\mathbf{k})
\frac{\mathbf{k}\cdot\mathbf{q}_1}{q_1^2}
\sum_{m=1}^{\infty}\sum_{l=1}^{\infty}\dot{a}a^{m+l-1}
h_m(\mathbf{q}_1,\tau)g_l(\mathbf{q}_2,\tau)
\theta_{m,c}(\mathbf{q}_1)\delta_{l,c}(\mathbf{q}_2),
\\
&&\sum_{n=1}^{\infty}\left[
\left(\ddot{a}(\tau)a^{n-1}(\tau)+\dot{a}^2(\tau)a^{n-2}(\tau)(n-1)\right)h_n(\mathbf{k},\tau)\theta_{n,c}(\mathbf{k})
+\dot{a}(\tau)a^{n-1}(\tau)\dot{h}_n(\mathbf{k},\tau)\theta_{n,c}(\mathbf{k})
\right.
\nonumber\\
&&+\left.\frac{2}{\tau}\dot{a}(\tau)a^{n-1}(\tau)h_n(\mathbf{k})(\mathbf{k},\tau)\theta_{n,c}(\mathbf{k})
+\frac{6}{\tau^2}a^n(\tau)\left(f_c+f_bg_n(\mathbf{k},\tau)\right)\delta_{n,c}(\mathbf{k})
\right]
\nonumber\\
&&=-\frac1{(2\pi)^3}\int\!\int d\mathbf{q}_1d\mathbf{q}_2
\delta_D(\mathbf{q}_1+\mathbf{q}_2-\mathbf{k})
\frac{k^2(\mathbf{q}_1\cdot\mathbf{q}_2)}{2q_1^2q_2^2}
\sum_{m=1}^{\infty}\sum_{l=1}^{\infty}\dot{a}^2(\tau)a^{m+l-2}(\tau)
h_m(\mathbf{q}_1,\tau)h_l(\mathbf{q}_2,\tau)
\theta_{m,c}(\mathbf{q}_1)\theta_{l,c}(\mathbf{q}_2)
\nonumber\\
&&+k^2c_s^2(\tau)\sum_{n=1}^{\infty}a^n(\tau)g_n(\mathbf{k},\tau)\delta_{n,c}(\mathbf{k})
\nonumber\\
&&-\frac1{2(2\pi)^3}k^2c_s^2(\tau)\int\!\int d\mathbf{q}_1d\mathbf{q}_2
\delta_D(\mathbf{q}_1+\mathbf{q}_2-\mathbf{k})
\sum_{m=1}^{\infty}\sum_{l=1}^{\infty}a^{m+l}(\tau)
g_m(\mathbf{q}_1,\tau)g_l(\mathbf{q}_2,\tau)
\delta_{m,c}(\mathbf{q}_1)\delta_{l,c}(\mathbf{q}_2)
\nonumber\\
&&+\frac1{3(2\pi)^6}k^2c_s^2(\tau)\int\!\int\!\int d\mathbf{q}_1
d\mathbf{q}_2d\mathbf{q}_3
\delta_D(\mathbf{q}_1+\mathbf{q}_2+\mathbf{q}_3-\mathbf{k})
\nonumber\\
&&\times\sum_{m=1}^{\infty}\sum_{l=1}^{\infty}\sum_{p=1}^{\infty}a^{m+l+p}(\tau)
g_m(\mathbf{q}_1,\tau)g_l(\mathbf{q}_2,\tau)g_p(\mathbf{q}_3,\tau)
\delta_{m,c}(\mathbf{q}_1)\delta_{l,c}(\mathbf{q}_2)
\delta_{p,c}(\mathbf{q}_3).
\end{eqnarray}
From now on, we shall write the sound speed, $c_s$, in terms of the
usual Jeans wavenumber, $k_J$, as $c_s=\sqrt{6}/(k_J\tau)$. 
We shall ignore the inhomogeneity in $c_s$ (i.e., spatial dependence of
$c_s$) throughout this paper. For the linear analysis for $\nabla c_s\ne
0$, see \citet{naoz/barkana:2005}.

\subsection{Second Order Solutions}

We have derived the linear filtering function, $g_1(k)$, in
Eq.~[\ref{eq:g(0)_1}]. For $n=2$, the continuity and Euler equations are
given by
\begin{eqnarray}
&&\delta_{2,c}(\mathbf{k})\dot{g}_2(\mathbf{k},\tau)+\frac{4}{\tau}\delta_{2,c}(\mathbf{k})g_2(\mathbf{k},\tau)
+\frac{2}{\tau}\theta_{2,c}(\mathbf{k})h_2(\mathbf{k},\tau)
\nonumber
\\
&&=\frac{2}{\tau}\frac1{(2\pi)^3}
\int\!\int d\mathbf{q}_1d\mathbf{q}_2
\delta_D(\mathbf{q}_1+\mathbf{q}_2-\mathbf{k})
\frac{\mathbf{k}\cdot\mathbf{q}_1}{q_1^2}
\delta_{1,c}(\mathbf{q}_1)\delta_{1,c}(\mathbf{q}_2)
g_1(\mathbf{q}_1)g_1(\mathbf{q}_2)
\nonumber
\\
&&\equiv\frac{2}{\tau}A_2(\mathbf{k}),
\label{eq:continuity_n2_app}\\
&&\frac{10}{\tau^2}\theta_{2,c}(\mathbf{k})h_2(\mathbf{k},\tau)
+\frac{2}{\tau}\theta_{2,c}(\mathbf{k})\dot{h}_2(\mathbf{k},\tau)
+\frac{6}{\tau^2}\delta_{2,c}(\mathbf{k})
-\frac{6}{\tau^2}\frac{k^2}{k_J^2}\delta_{2,c}(\mathbf{k})g_2(\mathbf{k},\tau)
\nonumber
\\
&&=\frac{4}{\tau^2}\frac1{(2\pi)^3}
\int\!\int d\mathbf{q}_1d\mathbf{q}_2
\delta_D(\mathbf{q}_1+\mathbf{q}_2-\mathbf{k})
\left[
-\frac{3}{4}\frac{k^2}{k_J^2}
-\frac{k^2(\mathbf{q}_1\cdot\mathbf{q}_2)}{2q_1^2q_2^2}
\right]
\delta_{1,c}(\mathbf{q}_1)\delta_{1,c}(\mathbf{q}_2)g_1(\mathbf{q}_1)g_1(\mathbf{q}_2)
\nonumber
\\
&&\equiv\frac{4}{\tau^2}B_2(\mathbf{k}).
\label{eq:euler_n2_app}
\end{eqnarray}
Here, $\theta_{1,c}(\mathbf{k})=-\delta_{1,c}(\mathbf{k})$.
Combining Eqs.~[\ref{eq:continuity_n2_app}] and [\ref{eq:euler_n2_app}], we get
the second order inhomogeneous partial differential equation:

\begin{eqnarray}
\ddot{g}_2(\mathbf{k},\tau)+\frac{10}{\tau^2}\dot{g}_2(\mathbf{k},\tau)
+\frac{1}{\tau^2}\left[20+6\frac{k^2}{k_J^2}\right]g_2(\mathbf{k},\tau)
+\frac{1}{\tau^2}\left[-6-\frac{10A_2(\mathbf{k})}{\delta_{2,c}(\mathbf{k})}
+\frac{4B_2(\mathbf{k})}{\delta_{2,c}(\mathbf{k})}\right]=0,
\label{eq:deq_g2_app}
\end{eqnarray}
where $\delta_{2,c}(\mathbf{k})$ is given by
\begin{equation}
\delta_{2,c}(\mathbf{k})=\frac{1}{(2\pi)^3}
\int d\mathbf{q}F_2^{(s)}(\mathbf{q},\mathbf{k}-\mathbf{q})
\delta_{1,c}(\mathbf{q})\delta_{1,c}(\mathbf{k}-\mathbf{q}).
\end{equation}
Solving the above differential equation, we have:
\begin{equation}
g_2(\mathbf{k},\tau)=\frac{6
+\frac{10A_2(\mathbf{k})}{\delta_{2,c}(\mathbf{k})}
-\frac{4B_2(\mathbf{k})}{\delta_{2,c}(\mathbf{k})}}
{20+6\frac{k^2}{k_J^2}}
+\mathcal{O}(\tau^{-9/2}),
\label{eq:g2_before_simplify_app}
\end{equation}
where the oscillation component,
\begin{equation}
\mathcal{O}(\tau^{-9/2})\propto\tau^{-\frac{9}{2}\left(
1\pm\sqrt{1-\frac4{81}(20+6\frac{k^2}{k_J^2})}
\right)},
\end{equation}
decays for any choice of $0\le k/k_J$. The 2nd-order filtering function
for the velocity divergence field, $h_2({\mathbf h},\tau)$, is given by
\begin{equation}
h_2(\mathbf{k})=\frac{1}{\theta_{2,c}(\mathbf{k})}
\left[A_2(\mathbf{k})-2\delta_{2,c}(\mathbf{k})g_2(\mathbf{k})
\right],
\end{equation}
where we have ignored the decaying term.

Using the explicit forms of $A_2({\mathbf k})$ and $B_2({\mathbf k})$
given by Eqs.~[\ref{eq:continuity_n2_app}] and [\ref{eq:euler_n2_app}],
respectively, we obtain
\begin{equation}
g_2(\mathbf{k},\tau)=\frac{\frac{10}3
-\frac73\left[1-\frac{\delta_{2,c}'(\mathbf{k})}{\delta_{2,c}(\mathbf{k})}
\right]
}
{\frac{10}3+\frac{k^2}{k_J^2}}
+\mathcal{O}(\tau^{-9/2}),
\label{eq:g2_after_simplify_app}
\end{equation}
where $\delta_{2,c}'$ is 
\begin{eqnarray}
\delta'_{2,c}(\mathbf{k})=
\frac{1}{(2\pi)^3}
\int d\mathbf{q}
\mathcal{F}_2^{(s)}(\mathbf{q},\mathbf{k}-\mathbf{q})
\delta_{1,c}(\mathbf{q})\delta_{1,c}(\mathbf{k}-\mathbf{q}),
\end{eqnarray}
where
\begin{equation}
\mathcal{F}_2^{(s)}(\mathbf{q}_1,\mathbf{q}_2)\equiv
\left[
F_2^{(s)}(\mathbf{q}_1,\mathbf{q}_2)
+\frac3{14}\frac{k^2}{k^2_J}\right]
g_1(\mathbf{q}_1)g_1(\mathbf{q}_2).
\end{equation}
In the limit where $k_J\to\infty$,
$\mathcal{F}_2^{(s)}(\mathbf{q}_1,\mathbf{q}_2)=F_2^{(s)}(\mathbf{q}_1,\mathbf{q}_2)$,
and thus $g_2\to1$. For the velocity divergence filtering function, we find
\begin{eqnarray}
h_2(\mathbf{k})\!&=&\!\frac1{\theta_{2,c}(\mathbf{k})}\!
\left[
\frac1{(2\pi)^3}\!\int\!\!\!\int\!
d\mathbf{q}_1d\mathbf{q}_2
\delta_D(\mathbf{q}_1+\mathbf{q}_2-\mathbf{k})
\delta_{1,c}(\mathbf{q}_1)\delta_{1,c}(\mathbf{q}_2)
(2F_2^{(s)}\!(\mathbf{q}_1,\!\mathbf{q}_2)\!-G_2^{(s)}\!(\mathbf{q}_1,\!\mathbf{q}_2))
g_1(\mathbf{q}_1)g_1(\mathbf{q}_2)\!-\!2\delta_{2,c}(\mathbf{k})g_2(\mathbf{k})\right]\nonumber\\
&=&
\!\frac1{\theta_{2,c}(\mathbf{k})}\!
\left[
\frac1{(2\pi)^3}\!\int\!\!\!\int\!
d\mathbf{q}_1d\mathbf{q}_2
\delta_D(\mathbf{q}_1+\mathbf{q}_2-\mathbf{k})
\delta_{1,c}(\mathbf{q}_1)\delta_{1,c}(\mathbf{q}_2)\!
\left(\!1\!+\!\frac{(\mathbf{q}_1\cdot\mathbf{q}_2)(q_1^2+q_2^2)}{2q_1^2q_2^2}
\right)\!g_1(\mathbf{q}_1)g_1(\mathbf{q}_2)\right]
\!\!-\!2\frac{\delta_{2,c}(\mathbf{k})}{\theta_{2,c}(\mathbf{k})}g_2(\mathbf{k}),
\label{eq:h2_to_gn_app}
\end{eqnarray}
where we have used
$2F_2(\mathbf{q}_1,\mathbf{q}_2)-G_2(\mathbf{q}_1,\mathbf{q}_2)=\frac{\mathbf{k}\cdot\mathbf{q}_1}{q_1^2}$. 
This expression also converges to $h_2=1$ as we take the limit of $\mathbf{k}_J\to \infty$.

\subsection{Third Order Solutions}
For $n=3$, the continuity and Euler equations are given by
\begin{eqnarray}
&&3\dot{a}(\tau)a^2(\tau)g_3(\mathbf{k},\tau)\delta_{3,c}(\mathbf{k})
+a^3(\tau)\dot{g}_3(\mathbf{k},\tau)\delta_{3,c}(\mathbf{k})
+\dot{a}(\tau)a^2(\tau)h_3(\mathbf{k},\tau)\theta_{3,c}(\mathbf{k})
\nonumber\\
&&=\dot{a}(\tau)a^2(\tau)\frac1{(2\pi)^6}\int\!\int\!\int
d\mathbf{q}_1d\mathbf{q}_2d\mathbf{q}_3
\delta_D(\mathbf{q}_1+\mathbf{q}_2+\mathbf{q}_3-\mathbf{k})
\delta_{1,c}(\mathbf{q}_1)\delta_{1,c}(\mathbf{q}_2)
\delta_{1,c}(\mathbf{q}_3)
\nonumber\\
&&\times\left[
\frac{\mathbf{k}\cdot\mathbf{q}_1}{q_1^2}
g_1(\mathbf{q}_1)g_2(\mathbf{q}_{23})
F_2^{(s)}(\mathbf{q}_2,\mathbf{q}_3)
+\frac{\mathbf{k}\cdot\mathbf{q}_{12}}{q_{12}^2}
h_2(\mathbf{q}_{12})g_1(\mathbf{q}_3)
G_2^{(s)}(\mathbf{q}_1,\mathbf{q}_2)
\right]
\nonumber\\
&&\equiv \dot{a}(\tau)a^2(\tau)A_3(\mathbf{k}),
\\
&&\left[\ddot{a}(\tau)a^2(\tau)+2\dot{a}^2(\tau)a(\tau)\right]
h_3(\mathbf{k},\tau)\theta_{3,c}(\mathbf{k})
+\dot{a}(\tau)a^2(\tau)\dot{h}_3(\mathbf{k},\tau)\theta_{3,c}(\mathbf{k})
+\frac{2}{\tau}\dot{a}(\tau)a^2(\tau)h_3(\mathbf{k},\tau)\theta_{3,c}(\mathbf{k})
\nonumber\\
&&+\frac{6}{\tau^2}a^3(\tau)\delta_{3,c}(\mathbf{k})
-\frac{6}{\tau^2}\frac{k^2}{k^2_J}a^3(\tau)\delta_{3,c}(\mathbf{k})
\nonumber\\
&&=\dot{a}^2(\tau)a(\tau)\frac1{(2\pi)^6}\int\!\int\!\int d\mathbf{q}_1
d\mathbf{q}_2d\mathbf{q}_3
\delta_D(\mathbf{q}_1+\mathbf{q}_2+\mathbf{q}_3-\mathbf{k})
\delta_{1,c}(\mathbf{q}_1)\delta_{1,c}(\mathbf{q}_2)\delta_{1,c}(\mathbf{q}_3)
\nonumber\\
&&\times\left[
-\frac{k^2(\mathbf{q}_1\cdot\mathbf{q}_{23})}{2q_1^2q_{23}^2}
g_1(\mathbf{q}_1)h_2(\mathbf{q}_{23})
G_2^{(s)}(\mathbf{q}_2,\mathbf{q}_3)
-\frac{k^2(\mathbf{q}_{12}\cdot\mathbf{q}_3)}{2q_{12}^2q_3^2}
h_2(\mathbf{q}_{12})g_1(\mathbf{q}_3)
G_2^{(s)}(\mathbf{q}_1,\mathbf{q}_2)
\right.
\nonumber\\
&&-\left.\frac34\frac{k^2}{k_J^2}
g_1(\mathbf{q}_1)g_2(\mathbf{q}_{23})
F_2^{(s)}(\mathbf{q}_2,\mathbf{q}_3)
-\frac34\frac{k^2}{k_J^2}
g_2(\mathbf{q}_{12})g_1(\mathbf{q}_3)
F_2^{(s)}(\mathbf{q}_1,\mathbf{q}_2)
+\frac12\frac{k^2}{k_J^2}
g_1(\mathbf{q}_1)g_1(\mathbf{q}_2)g_1(\mathbf{q}_3)
\right]
\nonumber\\
&&\equiv\dot{a}^2(\tau)a(\tau)B_3(\mathbf{k}).
\end{eqnarray}

In an EdS universe, $a(\tau)=\frac{\tau^2}{9}$, we have
\begin{eqnarray}
\delta_{3,c}(\mathbf{k})\dot{g}_3(\mathbf{k},\tau)+\frac{6}{\tau}\delta_{3,c}(\mathbf{k})g_3(\mathbf{k},\tau)
+\frac{2}{\tau}\theta_{3,c}(\mathbf{k})h_3(\mathbf{k},\tau)
&=&\frac{2}{\tau}A_3(\mathbf{k}),
\label{eq:continuity_n3_app}
\\
\frac{14}{\tau^2}h_3(\mathbf{k},\tau)\theta_{3,c}(\mathbf{k})+\frac{2}{\tau}\dot{h}_3(\mathbf{k},\tau)\theta_{3,c}(\mathbf{k})
+\frac{6}{\tau^2}\delta_{3,c}(\mathbf{k})
-\frac{6}{\tau^2}\frac{k^2}{k_J^2}\delta_{3,c}(\mathbf{k})g_3(\mathbf{k},\tau)
&=&\frac{4}{\tau^2}B_3(\mathbf{k}).
\label{eq:euler_n3_app}
\end{eqnarray}
Combining Eqs.~[\ref{eq:continuity_n3_app}] and [\ref{eq:euler_n3_app}],
we have the second-order differential equation:
\begin{eqnarray}
\ddot{g}_3(\mathbf{k},\tau)+\frac{14}{\tau}\dot{g}_3(\mathbf{k},\tau)+\frac{1}{\tau^2}
\left(42+6\frac{k^2}{k^2_J}\right)g_3(\mathbf{k},\tau)
+\frac{1}{\tau^2}\left(
-6-\frac{14A_3(\mathbf{k})}{\delta_{3,c}(\mathbf{k})}+\frac{4B_3(\mathbf{k})}{\delta_{3,c}(\mathbf{k})}
\right)=0.
\label{eq:deq_g3_app}
\end{eqnarray}
Solving this, we obtain
\begin{equation}
g_3(\mathbf{k},\tau)=
\frac{1+\frac{7A_3(\mathbf{k})}{3\delta_{3,c}(\mathbf{k})}-\frac{2B_3(\mathbf{k})}{3\delta_{3,c}(\mathbf{k})}}
{7+\frac{k^2}{k^2_J}}
+\mathcal{O}(\tau^{-13/2}),
\label{eq:g3_app}
\end{equation}
where the oscillation component,
\begin{equation}
\mathcal{O}(\tau^{-13/2})\propto
\tau^{-13/2\left(1\pm\sqrt{1-\frac{24}{169}(7+\frac{k^2}{k^2_J})}\right)},
\end{equation}
decays for any $0\le \frac{k}{k_J}$. The velocity divergence filtering
function at the 3rd-order is
\begin{equation}
h_3(\mathbf{k})=\frac{1}{\theta_{3,c}(\mathbf{k})}
\left[A_3(\mathbf{k})-3\delta_{3,c}(\mathbf{k})g_3(\mathbf{k})
\right], 
\end{equation}
where we have ignored the decaying term.

Let us rewrite $7A_3({\mathbf k})-2B_3({\mathbf k})$ in
Eq.~[\ref{eq:g3_app}] as
\begin{eqnarray}
7A_3(\mathbf{k})-2B_3(\mathbf{k})&=&
\frac1{(2\pi)^6}\int\!\int\!\int
d\mathbf{q}_1d\mathbf{q}_2d\mathbf{q}_3
\delta_D(\mathbf{q}_1+\mathbf{q}_2+\mathbf{q}_3-\mathbf{k})
\delta_{1,c}(\mathbf{q}_1)\delta_{1,c}(\mathbf{q}_2)
\delta_{1,c}(\mathbf{q}_3)
\nonumber
\\
&&\times\left[
\frac{7\mathbf{k}\cdot\mathbf{q}_1}{q_1^2}
g_1(\mathbf{q}_1)g_2(\mathbf{q}_{23})
F_2^{(s)}(\mathbf{q}_2,\mathbf{q}_3)
+\frac{7\mathbf{k}\cdot\mathbf{q}_{12}}{q_{12}^2}
h_2(\mathbf{q}_{12})g_1(\mathbf{q}_3)
G_2^{(s)}(\mathbf{q}_1,\mathbf{q}_2)
\right.
\nonumber
\\
&&+\frac{k^2(\mathbf{q}_1\cdot\mathbf{q}_{23})}{q_1^2q_{23}^2}
g_1(\mathbf{q}_1)h_2(\mathbf{q}_{23})
G_2^{(s)}(\mathbf{q}_2,\mathbf{q}_3)
+\frac{k^2(\mathbf{q}_{12}\cdot\mathbf{q}_3)}{q_{12}^2q_3^2}
h_2(\mathbf{q}_{12})g_1(\mathbf{q}_3)
G_2^{(s)}(\mathbf{q}_1,\mathbf{q}_2)
\nonumber
\\
&&+\left.\frac32\frac{k^2}{k_J^2}
g_1(\mathbf{q}_1)g_2(\mathbf{q}_{23})
F_2^{(s)}(\mathbf{q}_2,\mathbf{q}_3)
+\frac32\frac{k^2}{k_J^2}
g_2(\mathbf{q}_{12})g_1(\mathbf{q}_3)
F_2^{(s)}(\mathbf{q}_1,\mathbf{q}_2)
-\frac{k^2}{k_J^2}
g_1(\mathbf{q}_1)g_1(\mathbf{q}_2)g_1(\mathbf{q}_3)
\right]
\nonumber
\\
&\equiv&\frac{18}{(2\pi)^6}\int\!\int\!\int
d\mathbf{q}_1d\mathbf{q}_2d\mathbf{q}_3
\delta_D(\mathbf{q}_1+\mathbf{q}_2+\mathbf{q}_3-\mathbf{k})
\mathcal{F}_3(\mathbf{q}_1,\mathbf{q}_2,\mathbf{q}_3)
\delta_{1,c}(\mathbf{q}_1)\delta_{1,c}(\mathbf{q}_2)
\delta_{1,c}(\mathbf{q}_3)
\nonumber
\\
&\equiv& 18\delta'_{3,c}(\mathbf{k}).
\end{eqnarray}
The new kernel, ${\cal F}_3(\mathbf{q}_1,\mathbf{q}_2,\mathbf{q}_3)$,
can be symmetrized as
\begin{eqnarray}
&&\mathcal{F}^{(s)}_3(\mathbf{q}_1,\mathbf{q}_2,\mathbf{q}_3)
\nonumber\\
&&=\frac16\left[
\mathcal{F}_3(\mathbf{q}_1,\mathbf{q}_2,\mathbf{q}_3)
+\mathcal{F}_3(\mathbf{q}_1,\mathbf{q}_3,\mathbf{q}_2)
+\mathcal{F}_3(\mathbf{q}_2,\mathbf{q}_1,\mathbf{q}_3)
+\mathcal{F}_3(\mathbf{q}_2,\mathbf{q}_3,\mathbf{q}_1)
+\mathcal{F}_3(\mathbf{q}_3,\mathbf{q}_1,\mathbf{q}_2)
+\mathcal{F}_3(\mathbf{q}_3,\mathbf{q}_2,\mathbf{q}_1)
\right]
\nonumber
\\
&&=
\frac7{54}\mathbf{k}\cdot\left[
F_2^{(s)}(\mathbf{q}_2,\mathbf{q}_3)\frac{\mathbf{q}_1}{q_1^2}
g_1(\mathbf{q}_1)g_2(\mathbf{q}_{23})
+F_2^{(s)}(\mathbf{q}_1,\mathbf{q}_3)\frac{\mathbf{q}_2}{q_2^2}
g_1(\mathbf{q}_2)g_2(\mathbf{q}_{13})
+F_2^{(s)}(\mathbf{q}_1,\mathbf{q}_2)\frac{\mathbf{q}_3}{q_3^2}
g_1(\mathbf{q}_3)g_2(\mathbf{q}_{12})
\right]
\nonumber
\\
&&+\frac1{27}k^2\left[
G_2^{(s)}(\mathbf{q}_2,\mathbf{q}_3)\frac{\mathbf{q}_1\cdot\mathbf{q}_{23}}{q_1^2q_{23}^2}
g_1(\mathbf{q}_1)h_2(\mathbf{q}_{23})
+G_2^{(s)}(\mathbf{q}_1,\mathbf{q}_3)\frac{\mathbf{q}_2\cdot\mathbf{q}_{13}}{q_2^2q_{13}^2}
g_1(\mathbf{q}_2)h_2(\mathbf{q}_{13})
+G_2^{(s)}(\mathbf{q}_1,\mathbf{q}_2)\frac{\mathbf{q}_3\cdot\mathbf{q}_{12}}{q_3^2q_{12}^2}
g_1(\mathbf{q}_3)h_2(\mathbf{q}_{12})
\right]
\nonumber
\\
&&+\frac7{54}\mathbf{k}\cdot\left[
G_2^{(s)}(\mathbf{q}_2,\mathbf{q}_3)\frac{\mathbf{q}_{23}}{q_{23}^2}
g_1(\mathbf{q}_1)h_2(\mathbf{q}_{23})
+G_2^{(s)}(\mathbf{q}_1,\mathbf{q}_3)\frac{\mathbf{q}_{13}}{q_{13}^2}
g_1(\mathbf{q}_2)h_2(\mathbf{q}_{13})
+G_2^{(s)}(\mathbf{q}_1,\mathbf{q}_2)\frac{\mathbf{q}_{12}}{q_{12}^2}
g_1(\mathbf{q}_3)h_2(\mathbf{q}_{12})
\right]
\nonumber
\\
&&+\frac1{18}\frac{k^2}{k^2_J}\left[
g_1(\mathbf{q}_1)g_2(\mathbf{q}_{23})
F_2^{(s)}(\mathbf{q}_2,\mathbf{q}_3)
+g_1(\mathbf{q}_2)g_2(\mathbf{q}_{13})
F_2^{(s)}(\mathbf{q}_1,\mathbf{q}_3)
+g_1(\mathbf{q}_3)g_2(\mathbf{q}_{12})
F_2^{(s)}(\mathbf{q}_1,\mathbf{q}_2)
-g_1(\mathbf{q}_1)g_1(\mathbf{q}_2)g_1(\mathbf{q}_3)
\right].\nonumber\\
\label{eq:f3prims_app}
\end{eqnarray}
In the limit of $k_J\to\infty$, $\mathcal{F}_3\to F_3$, and $g_3(k)=1$.
Using $\delta'_{3,c}({\mathbf k})$ introduced above, 
we write $g_3$ as
\begin{equation}
g_3(\mathbf{k})=\frac{7-6\left[1-\frac{\delta_{3,c}'(\mathbf{k})}{\delta_{3,c}(\mathbf{k})}\right]}
{7+\frac{k^2}{k^2_J}}.
\end{equation}

\section{3PT Total Power Spectrum}
\label{sec:3pt_power_app}

\begin{figure}[t]
  \includegraphics[width=18cm]{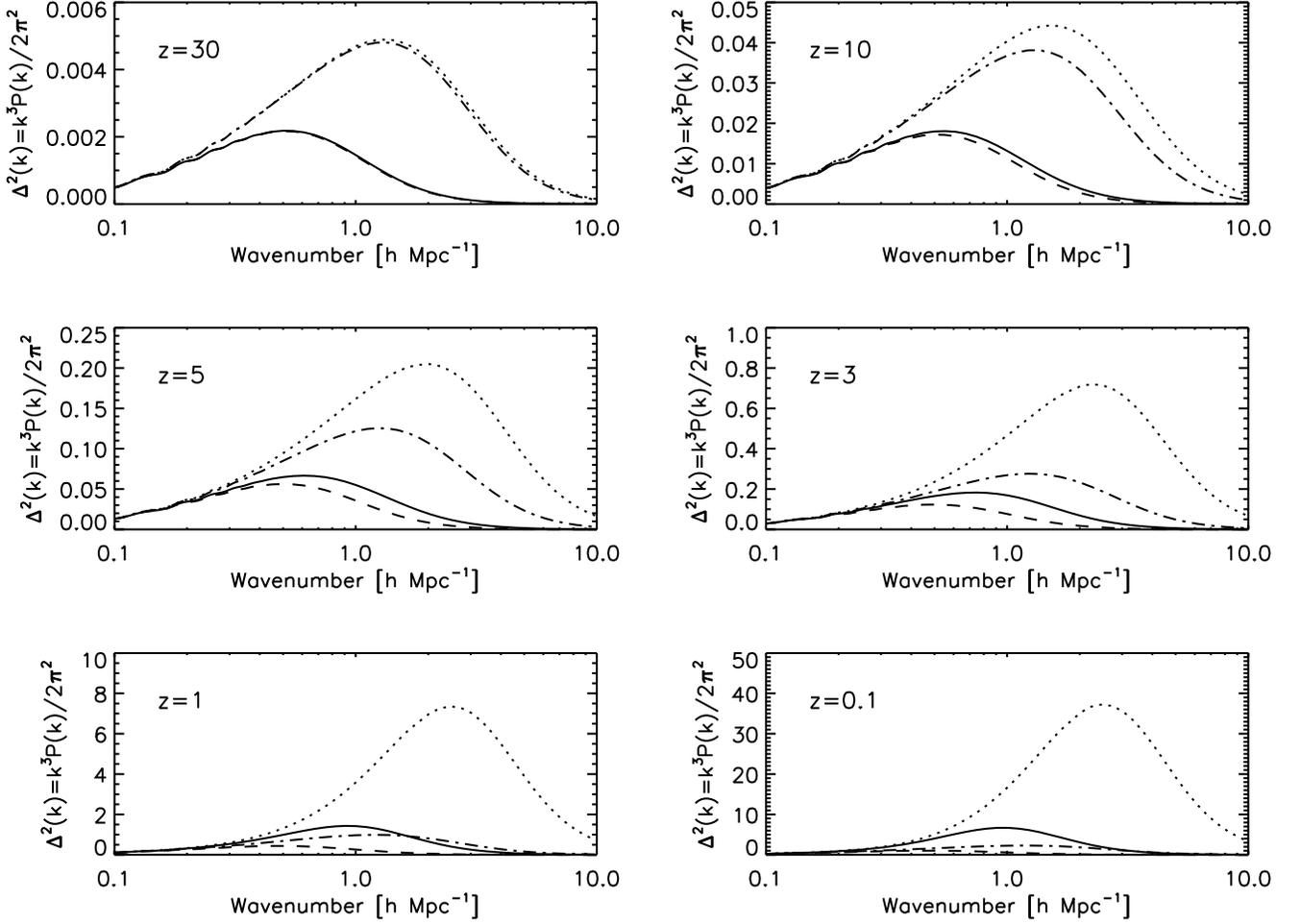}
\label{fig:3pt_dimless_b}
\caption{%
The dimensionless power spectra, $\Delta^2(k)\equiv k^3P(k)/(2\pi^2)$, 
for a matter component
with pressure (i.e., baryon, neutrino, etc) are shown for several
redshifts ($z=0.1$, $1.0$, $3.0$, $5.0$, $10$ and $30$).
We show the non-linear calculations with 3PT 
in the solid and dotted lines for  $k_J=1.0$ and $3.0~h~{\rm Mpc^{-1}}$,
 respectively.
We also show the linear calculations 
in the dashed and dot-dashed lines
for  $k_J=1.0$ and $3.0~h~{\rm Mpc^{-1}}$,
 respectively.
}%
\end{figure}
We calculate the power spectrum of the total matter fluctuations,
$\delta=f_c\delta_c+f_b\delta_b=f_c\delta_c+(1-f_c)\delta_b$, which is given, up to the third-order
in perturbations, by 
\begin{eqnarray}
\delta(\mathbf{k},\tau)&=&f_c\delta_c(\mathbf{k},\tau)+f_b\delta_b(\mathbf{k},\tau)
\nonumber
\\
&=&f_c\left[\delta_{1,c}(\mathbf{k},\tau)+\delta_{2,c}(\mathbf{k},\tau)+\delta_{3,c}(\mathbf{k},\tau)\right]
+(1-f_c)\left[\delta_{1,b}(\mathbf{k},\tau)+\delta_{2,b}(\mathbf{k},\tau)+\delta_{3,b}(\mathbf{k},\tau)\right]
\nonumber
\\
&=&f_c\left[\delta_{1,c}(\mathbf{k},\tau)+\delta_{2,c}(\mathbf{k},\tau)+\delta_{3,c}(\mathbf{k},\tau)\right]
+(1-f_c)\left[g_1(k)\delta_{1,c}(\mathbf{k},\tau)+g_2(\mathbf{k})\delta_{2,c}(\mathbf{k},\tau)+g_3(\mathbf{k})\delta_{3,c}(\mathbf{k},\tau)\right].
\end{eqnarray}
The power spectrum is
\begin{eqnarray}
(2\pi)^3P_{tot}(k)\delta_D(\mathbf{k}+\mathbf{k}')
&=&\langle\delta(\mathbf{k})\delta(\mathbf{k}')\rangle
\nonumber
\\
&=&\langle\{f_c\delta_c(\mathbf{k})+(1-f_c)\delta_b(\mathbf{k})\}
\{f_c\delta_c(\mathbf{k}')+(1-f_c)\delta_b(\mathbf{k}')\}\rangle
\nonumber
\\
&=&f_c^2\langle\delta_c(\mathbf{k})\delta_c(\mathbf{k}')\rangle
+2f_c(1-f_c)\langle\delta_b(\mathbf{k})\delta_c(\mathbf{k}')\rangle
+(1-f_c)^2\langle\delta_b(\mathbf{k})\delta_b(\mathbf{k}')\rangle
\nonumber
\\
&\equiv& (2\pi)^3\left[
f_c^2P_c(k)+2f_c(1-f_c)P_{b,c}(k)+(1-f_c)^2P_b(k)
\right]\delta_D(\mathbf{k}+\mathbf{k}'),
\end{eqnarray}
where $P_c$, $P_{b,c}$ and $P_b$ are
\begin{eqnarray}
(2\pi)^3P_{c}(k)\delta_D(\mathbf{k}+\mathbf{k}')
&=&\langle\delta_c(\mathbf{k})\delta_c(\mathbf{k}')\rangle
\nonumber
\\
&=&\langle\{\delta_{1,c}(\mathbf{k})
+\delta_{2,c}(\mathbf{k})
+\delta_{3,c}(\mathbf{k})\}
\{\delta_{1,c}(\mathbf{k}')
+\delta_{2,c}(\mathbf{k}')
+\delta_{3,c}(\mathbf{k}')\}\rangle
\nonumber
\\
&=&\langle\delta_{1,c}(\mathbf{k})\delta_{1,c}(\mathbf{k}')\rangle
+2\langle\delta_{1,c}(\mathbf{k})\delta_{3,c}(\mathbf{k}')\rangle
+\langle\delta_{2,c}(\mathbf{k})\delta_{2,c}(\mathbf{k}')\rangle\nonumber
\\
&\equiv&(2\pi)^3\left[P_{11,c}(k)+2P_{13,c}(k)+P_{22,c}(k)
\right]
\delta_D(\mathbf{k}+\mathbf{k}'),
\\
(2\pi)^3P_{bc}(k)\delta_D(\mathbf{k}+\mathbf{k}')
&=&\langle\delta_b(\mathbf{k})\delta_c(\mathbf{k}')\rangle
\nonumber
\\
&=&\langle\{g_1(k)\delta_{1,c}(\mathbf{k})
+g_2(\mathbf{k})\delta_{2,c}(\mathbf{k})
+g_3(\mathbf{k})\delta_{3,c}(\mathbf{k})\}
\{\delta_{1,c}(\mathbf{k}')
+\delta_{2,c}(\mathbf{k}')
+\delta_{3,c}(\mathbf{k}')\}\rangle
\nonumber
\\
&=&g_1(k)\langle\delta_{1,c}(\mathbf{k})\delta_{1,c}(\mathbf{k}')\rangle
+g_1(k)\langle\delta_{1,c}(\mathbf{k})\delta_{3,c}(\mathbf{k}')\rangle
+\langle g_3(\mathbf{k})\delta_{3,c}(\mathbf{k})\delta_{1,c}(\mathbf{k}')\rangle
+\langle g_2(\mathbf{k})\delta_{2,c}(\mathbf{k})\delta_{2,c}(\mathbf{k}')\rangle\nonumber
\\
&\equiv&(2\pi)^3\left[P_{11,bc}(k)+2P_{13,bc}(k)+P_{22,bc}(k)
\right]
\delta_D(\mathbf{k}+\mathbf{k}'),
\\
(2\pi)^3P_{b}(k)\delta_D(\mathbf{k}+\mathbf{k}')
&=&\langle\delta_b(\mathbf{k})\delta_b(\mathbf{k}')\rangle
\nonumber
\\
&=&\langle\{g_1(k)\delta_{1,c}(\mathbf{k})
+g_2(\mathbf{k})\delta_{2,c}(\mathbf{k})
+g_3(\mathbf{k})\delta_{3,c}(\mathbf{k})\}
\times
\{g_1(k')\delta_{1,c}(\mathbf{k}')
+g_2(\mathbf{k}')\delta_{2,c}(\mathbf{k}')
+g_3(\mathbf{k}')\delta_{3,c}(\mathbf{k}')\}\rangle
\nonumber
\\
&=&g_1^2(k)\langle\delta_{1,c}(\mathbf{k})\delta_{1,c}(\mathbf{k}')\rangle
+2g_1(k)\langle\delta_{1,c}(\mathbf{k})g_3(\mathbf{k}')\delta_{3,c}(\mathbf{k}')\rangle
+\langle g_2(\mathbf{k})\delta_{2,c}(\mathbf{k})g_2(\mathbf{k}')\delta_{2,c}(\mathbf{k}')\rangle\nonumber
\\
&\equiv&(2\pi)^3\left[P_{11,b}(k)+2P_{13,b}(k)+P_{22,b}(k)
\right]
\delta_D(\mathbf{k}+\mathbf{k}'),
\end{eqnarray}
respectively.

Now, $P_{11,c}(k)$, $P_{13,c}(k)$ and $P_{22,c}(k)$
can be numerically calculated with the corresponding kernels,
$F_2^{(s)}$ and $F_3^{(s)}$;
\begin{eqnarray}
(2\pi)^3P_{11,bc}(k)\delta_D(\mathbf{k}+\mathbf{k}')
&=&\langle\delta_{1,b}(\mathbf{k})\delta_{1,c}(\mathbf{k}')\rangle
\nonumber
\\
&=&g_1(k)\langle\delta_{1,c}(\mathbf{k})\delta_{1,c}(\mathbf{k}')\rangle,
\\
(2\pi)^3P_{13,bc}(k)\delta_D(\mathbf{k}+\mathbf{k}')
&=&\frac12\left[\langle\delta_{1,b}(\mathbf{k})\delta_{3,c}(\mathbf{k}')\rangle
+\langle\delta_{1,c}(\mathbf{k})\delta_{3,b}(\mathbf{k}')\rangle\right]
\nonumber
\\
&=&\frac12\left[
g_1(k)\langle\delta_{1,c}(\mathbf{k})\delta_{3,c}(\mathbf{k}')\rangle
+\langle\delta_{1,c}(\mathbf{k})g_3(\mathbf{k}')\delta_{3,c}(\mathbf{k}')\rangle
\right]
\nonumber
\\
&=&\frac12\left[\left(g_1(k)+\frac{1}{7+\frac{k^2}{k^2_J}}\right)
\langle\delta_{1,c}(\mathbf{k})\delta_{3,c}(\mathbf{k}')\rangle
+\frac{6}{7+\frac{k^2}{k^2_J}}
\langle\delta_{1,c}(\mathbf{k})\delta'_{3,c}(\mathbf{k}')\rangle
\right],\label{eq:p13bc_app}
\\
(2\pi)^3P_{22,bc}(k)\delta_D(\mathbf{k}+\mathbf{k}')
&=&\langle\delta_{2,b}(\mathbf{k})\delta_{2,c}(\mathbf{k}')\rangle
\nonumber
\\
&=&\langle g_2(\mathbf{k})\delta_{2,c}(\mathbf{k})\delta_{2,c}(\mathbf{k}')\rangle
\nonumber
\\
&=&\frac{1}{\frac{10}3+\frac{k^2}{k^2_J}}
\left[\langle\delta_{2,c}(\mathbf{k})\delta_{2,c}(\mathbf{k}')\rangle
+\frac73\langle\delta'_{2,c}(\mathbf{k})\delta_{2,c}(\mathbf{k}')\rangle
\right],
\label{eq:p22bc_app}
\\
(2\pi)^3P_{11,b}(k)\delta_D(\mathbf{k}+\mathbf{k}')
&=&\langle\delta_{1,b}(\mathbf{k})\delta_{1,b}(\mathbf{k}')\rangle
\nonumber
\\
&=&g_1^2(k)\langle\delta_{1,c}(\mathbf{k})\delta_{1,c}(\mathbf{k}')\rangle,
\\
(2\pi)^3P_{13,b}(k)\delta_D(\mathbf{k}+\mathbf{k}')
&=&\langle\delta_{1,b}(\mathbf{k})\delta_{3,b}(\mathbf{k}')\rangle
\nonumber
\\
&=&g_1(k)\langle\delta_{1,c}(\mathbf{k})g_3(\mathbf{k}')\delta_{3,c}(\mathbf{k}')\rangle
\nonumber
\\
&=&\frac{g_1(k)}{7+\frac{k^2}{k^2_J}}
\left[\langle\delta_{1,c}(\mathbf{k})\delta_{3,c}(\mathbf{k}')\rangle
+6\langle\delta_{1,c}(\mathbf{k})\delta'_{3,c}(\mathbf{k}')\rangle
\right],
\label{eq:p13b_app}
\\
(2\pi)^3P_{22,b}(k)\delta_D(\mathbf{k}+\mathbf{k}')
&=&\langle\delta_{2,b}(\mathbf{k})\delta_{2,b}(\mathbf{k}')\rangle
\nonumber
\\
&=&\langle g_2(\mathbf{k})\delta_{2,c}(\mathbf{k})g_2(\mathbf{k}')\delta_{2,c}(\mathbf{k}')\rangle
\nonumber
\\
&=&\frac1{\left(\frac{10}3+\frac{k^2}{k^2_J}\right)^2}
\left[
\langle\delta_{2,c}(\mathbf{k})\delta_{2,c}(\mathbf{k}')\rangle
+\frac{14}3\langle\delta_{2,c}(\mathbf{k})\delta'_{2,c}(\mathbf{k}')\rangle
+\frac{49}9\langle\delta'_{2,c}(\mathbf{k})\delta'_{2,c}(\mathbf{k}')\rangle
\right].
\label{eq:p22b_app}
\end{eqnarray}
The ensemble averages of the products involving
$\delta'_{n,c}(\mathbf{k})$ are given by
\begin{eqnarray}
\langle\delta_{1,c}(\mathbf{k})\delta'_{3,c}(\mathbf{k}')\rangle
&=&3\delta_D(\mathbf{k}+\mathbf{k}')P_{11,c}(k)
\int d\mathbf{q}\mathcal{F}_3^{(s)}(\mathbf{q},-\mathbf{q},\mathbf{k})
P_{11,c}(q)\nonumber
\\
&=&6\pi\delta_D(\mathbf{k}+\mathbf{k}')P_{11,c}(k)
\int^{\infty}_0 dq\ q^2P_{11,c}(q)\int^1_{-1}d\mu
\mathcal{F}_3^{(s)}(\mathbf{q},-\mathbf{q},\mathbf{k}),
\label{eq:d1d3prime}
\\
\langle\delta_{2,c}(\mathbf{k})\delta'_{2,c}(\mathbf{k}')\rangle
&=&2\delta_D(\mathbf{k}+\mathbf{k}')\int d\mathbf{q}
P_{11,c}(q)P_{11,c}(|\mathbf{k}-\mathbf{q}|)
F_2^{(s)}(\mathbf{q},\mathbf{k}-\mathbf{q})
\mathcal{F}_2^{(s)}(\mathbf{q},\mathbf{k}-\mathbf{q}),
\label{eq:d2d2prime}
\\
\langle\delta'_{2,c}(\mathbf{k})\delta'_{2,c}(\mathbf{k}')\rangle
&=&2\delta_D(\mathbf{k}+\mathbf{k}')\int d\mathbf{q}
P_{11,c}(q)P_{11,c}(|\mathbf{k}-\mathbf{q}|)
\left[
\mathcal{F}_2^{(s)}(\mathbf{q},\mathbf{k}-\mathbf{q})
\right]^2.
\label{eq:d2primed2prime}
\end{eqnarray}
Here, the term, 
$\int
\frac{d\mathbf{q}}{(2\pi)^3}\mathcal{F}_3(\mathbf{q},-\mathbf{q},\mathbf{k})P_{11,c}(q)$,
in Eq.~[\ref{eq:d1d3prime}] is given by
\begin{eqnarray}
&&\int \frac{d\mathbf{q}}{(2\pi)^3}\mathcal{F}^{(s)}_3(\mathbf{q},-\mathbf{q},\mathbf{k})P_{11,c}(q)
\nonumber\\
&=&\int \frac{d\mathbf{q}}{(2\pi)^3}\left\{\frac7{54}\mathbf{k}\cdot\left[
F_2^{(s)}(-\mathbf{q},\mathbf{k})\frac{\mathbf{q}}{q^2}
g_1(\mathbf{q})g_2(\mathbf{k}-\mathbf{q})
-F_2^{(s)}(\mathbf{q},\mathbf{k})\frac{\mathbf{q}}{q^2}
g_1(\mathbf{q})g_2(\mathbf{k}+\mathbf{q})
\right]\right.
\nonumber
\\
&&+\frac2{27}k^2\left[
F_2^{(s)}(-\mathbf{q},\mathbf{k})\frac{\mathbf{q}\cdot(\mathbf{k}-\mathbf{q})}{q^2(\mathbf{k}-\mathbf{q})^2}
g_1(\mathbf{q})g_2(\mathbf{k}-\mathbf{q})
-F_2^{(s)}(\mathbf{q},\mathbf{k})\frac{\mathbf{q}\cdot(\mathbf{k}+\mathbf{q})}{q^2(\mathbf{k}+\mathbf{q})^2}
g_1(\mathbf{q})g_2(\mathbf{k}+\mathbf{q})
\right]
\nonumber
\\
&&+\frac{14}{54}\mathbf{k}\cdot\left[
F_2^{(s)}(-\mathbf{q},\mathbf{k})\frac{\mathbf{k}-\mathbf{q}}{(\mathbf{k}-\mathbf{q})^2}
g_1(\mathbf{q})g_2(\mathbf{k}-\mathbf{q})
+F_2^{(s)}(\mathbf{q},\mathbf{k})\frac{\mathbf{k}+\mathbf{q}}{(\mathbf{k}+\mathbf{q})^2}
g_1(\mathbf{q})g_2(\mathbf{k}+\mathbf{q})
\right]
\nonumber
\\
&&-\frac1{27}k^2\left[
\left(1+\frac{(-\mathbf{q}\cdot\mathbf{k})(q^2+k^2)}{2q^2k^2}
\right)
\frac{\mathbf{q}\cdot(\mathbf{k}-\mathbf{q})}
{q^2(\mathbf{k}-\mathbf{q})^2}
g_1^2(\mathbf{q})g_1(\mathbf{k})
-\left(1+\frac{(\mathbf{q}\cdot\mathbf{k})(q^2+k^2)}{2q^2k^2}
\right)
\frac{\mathbf{q}\cdot(\mathbf{k}+\mathbf{q})}
{q^2(\mathbf{k}+\mathbf{q})^2}
g_1^2(\mathbf{q})g_1(\mathbf{k})
\right]
\nonumber
\\
&&-\frac7{54}\mathbf{k}\cdot\left[
\left(1+\frac{(-\mathbf{q}\cdot\mathbf{k})(q^2+k^2)}{2q^2k^2}
\right)
\frac{\mathbf{k}-\mathbf{q}}
{(\mathbf{k}-\mathbf{q})^2}
g_1^2(\mathbf{q})g_1(\mathbf{k})
+\left(1+\frac{(\mathbf{q}\cdot\mathbf{k})(q^2+k^2)}{2q^2k^2}
\right)
\frac{\mathbf{k}+\mathbf{q}}
{(\mathbf{k}+\mathbf{q})^2}
g_1^2(\mathbf{q})g_1(\mathbf{k})
\right]
\nonumber
\\
&&+\left.\frac1{18}\frac{k^2}{k^2_J}\left[
g_1(\mathbf{q})g_2(\mathbf{k}-\mathbf{q})
F_2^{(s)}(-\mathbf{q},\mathbf{k})
+g_1(\mathbf{q})g_2(\mathbf{k}+\mathbf{q})
F_2^{(s)}(\mathbf{q},\mathbf{k})
-g_1^2(\mathbf{q})g_1(\mathbf{k})
\right]\right\}P_{11,c}(q),
\end{eqnarray}
where we have used Eq.~[\ref{eq:h2_to_gn_app}] and
$F_2^{(s)}(\mathbf{q},-\mathbf{q})=G_2^{(s)}(\mathbf{q},-\mathbf{q})=0$.
We then calculate the angular average of $\mathcal{F}_3^{(s)}$, i.e.,
$\int d\mu \mathcal{F}_3^{(s)}$, for  the linear filtering function of
$g_1(k)=1/(1+k^2/k_J^2)$: 
\begin{eqnarray}
&&\int^1_{-1}d\mu\mathcal{F}^{(s)}_3=
\frac1{612360r^8s(1+r^2)(r^2+s^2)^2}[\![30r^2s^3[-14000s^6+810r^{10}(1+s^2)
+900r^2s^4(-7+5s^2)
\nonumber\\
&&+60r^4s^2(105-125s^2+78s^4)+9r^8(321-248s^2+159s^4)+27r^6(126-87s^2+70s^4+9s^6)]
\nonumber\\
&&-243r^8(-7+5s^2+2s^4)[5(r^4+s^2)(-1+s^2)^2+r^2(5-5s^2-19s^4+5s^6)]\ln\frac{1+s}{|1-s|}
\nonumber\\
&&+[10s^2+3r^2(1+s^2)][-35s^2+3r^2(-7+s^2)]
[-2000s^6+135r^8(-1+s^2)^2
\nonumber\\
&&+240r^4s^2(3-4s^2+3s^4)+300r^2(s^4+s^6)+27r^6(5+5s^2-9s^4+5s^6)]
\frac12\ln\left[\frac{10s^2+3r^2(1+s)^2}{10s^2+3r^2(1-s)^2}\right]]\!],
\end{eqnarray}
where $r\equiv k/k_J$ and $s\equiv k/q$.
We find that the calculation of $\mathcal{F}_3$ is numerically unstable as
$k/k_J\to 0$ ($r\rightarrow 0$). The exact limit of $\mathcal{F}_3$ 
is $\lim_{k/k_J\to 0}\mathcal{F}_3\to F_3$, and thus one may replace
$\mathcal{F}_3$ with $F_3$ for a sufficiently small value of $k/k_J$.

Finally, we generalize the above results from an EdS universe to
general cosmological models, by writing 
\begin{equation}
 \frac{a^2(\tau)}{a^2(\tau_i)} P_{11}(k,\tau_i)
\to 
P_{11}(k,\tau)
= \frac{D^2(\tau)}{D^2(\tau_i)} 
\left(\frac{\delta_{1,c+}^{(1)}(k,\tau)/\delta_{1,c+}^{(0)}(k,\tau)}
{\delta_{1,c+}^{(1)}(k,\tau_*)/\delta_{1,c+}^{(0)}(k,\tau_*)}\right)^2
P_{11}(k,\tau_i),
\end{equation}
where $\tau_i$ is some arbitrary epoch, $\tau_*$ is the epoch
where the pressure effect becomes non-negligible (i.e., reionization
epoch for baryons and non-relativistic transition for massive neutrinos),
and $D(\tau)$ is the linear growth factor appropriate to a given
cosmological model.
We obtain Eq.~[\ref{eq:p22bc}] from combining
Eqs.~[\ref{eq:p22bc_app}], 
[\ref{eq:d2d2prime}], and $P_{22,c}$ given by Eq.~[\ref{eq:p22_app}].
Similarly, we obtain Eq.~[\ref{eq:p22b}] from combining
Eqs.~[\ref{eq:p22b_app}], 
[\ref{eq:d2primed2prime}], and $P_{22,c}$,
Eq.~[\ref{eq:p13bc}] from combining Eqs.~[\ref{eq:p13bc_app}], 
[\ref{eq:d1d3prime}], and $P_{13,c}$ given by Eq.~[\ref{eq:p13_app}], and
Eq.~[\ref{eq:p13b}] from combining Eqs.~[\ref{eq:p13b_app}], 
[\ref{eq:d1d3prime}], and $P_{13,c}$.

Figure \ref{fig:3pt_dimless_b} shows the dimensionless 3PT and linear
power spectra, $\Delta^2(k)=k^3P(k)/(2\pi^2)$, for a matter component with pressure at different redshifts
($z=0.1$, $1.0$, $3.0$, $5.0$, $10$ and $30$) with $k_J=1.0$ and
$3.0~h~{\rm Mpc^{-1}}$.
The 3PT and linear power spectra are similar
at the highest redshift, 
whereas the 3PT has significantly more power than the linear 
spectrum at larger wavenumbers as we go to lower redshifts. 
As a result, the filtering scale for a given linear filtering scale
migrates toward larger wavenumbers in lower redshifts.


\begin{thebibliography}{}
\bibitem[\protect\citeauthoryear{{Bernardeau} et~al.}{2002}]{bernardeau/etal:2002} Bernardeau, F., 
Colombi, S., Gazta{\~n}aga, E., \& Scoccimarro, R.\ 2002, \physrep, 367, 1 
\bibitem[\protect\citeauthoryear{{Gnedin}}{2000}]{gnedin:2000} Gnedin, N.~Y.\ 2000, \apj, 542, 
535 
\bibitem[\protect\citeauthoryear{{Gnedin} \& {Hui}}{1998}]{gnedin/hui:1998} Gnedin, N.~Y., \& Hui, L.\ 1998, \mnras, 296, 44 
\bibitem[\protect\citeauthoryear{{Hoeft} et~al.}{2006}]{hoeft/etal:2006} Hoeft, M., Yepes, G., 
Gottl{\"o}ber, S., \& Springel, V.\ 2006, \mnras, 371, 401 
\bibitem[\protect\citeauthoryear{{Hu} et~al.}{1998}]{hu/etal:1998} Hu, W., Eisenstein, D.~J., 
\& Tegmark, M.\ 1998, \prl, 80, 5255 
\bibitem[\protect\citeauthoryear{{Lesgourgues}
		      et~al.}{2009}]{lesgourgues/etal:2009} 
Lesgourgues, L., Matarrese, S., Pietroni, M., \& Riotto, A.\ 2009, 
arXiv:0901.4550
\bibitem[\protect\citeauthoryear{{Makino} et~al.}{1992}]{makino/etal:1992} Makino, N., Sasaki, M., 
\& Suto, Y.\ 1992, \prd, 46, 585 
\bibitem[\protect\citeauthoryear{{Naoz} \& {Barkana}}{2005}]{naoz/barkana:2005} Naoz, S., \& Barkana, R.\ 2005, \mnras, 362, 1047 
\bibitem[\protect\citeauthoryear{{Nusser}}{2000}]{nusser:2000} Nusser, A.\ 2000, \mnras, 317, 
902 
\bibitem[\protect\citeauthoryear{{Okamoto} et~al.}{2008}]{okamoto/etal:2008} Okamoto, T., Gao, L., 
\& Theuns, T.\ 2008, \mnras, 390, 920 
\bibitem[\protect\citeauthoryear{{Saito} et~al.}{{Saito} et~al.}{2008}]{saito/etal:2008} Saito, S., Takada, M., 
\& Taruya, A.\ 2008, \prl, 100, 191301 
\bibitem[\protect\citeauthoryear{{Takada} et~al.}{2006}]{takada/komatsu/futamase:2006} Takada, M., Komatsu, E., 
\& Futamase, T.\ 2006, \prd, 73, 083520 
\bibitem[\protect\citeauthoryear{{Takahashi}}{2008}]{takahashi:2008} Takahashi, R.\ 2008, 
Progress of Theoretical Physics, 120, 549 
\bibitem[\protect\citeauthoryear{{Zaldarriaga} et~al.}{2001}]{zaldarriaga/etal:2001} Zaldarriaga, M., 
Hui, L., \& Tegmark, M.\ 2001, \apj, 557, 519 
\bibitem[\protect\citeauthoryear{Weinberg}{Weinberg}{2008}]{weinberg:COS}
Weinberg, S. 2008, Cosmology (Oxford, UK: Oxford University Press)
\bibitem[\protect\citeauthoryear{{Wong}}{2008}]{wong:2008} Wong,
		Y.~Y.~Y.\ 2008, J. Cosmol. Astropart. Phys., 10, 035 
\end{thebibliography}
\end{document}